\documentclass[final,5p,times,twocolumn,anonymous]{elsarticle}

\usepackage{bbm} 
\usepackage{amssymb}
\usepackage{amsmath}
\usepackage{subcaption}
\usepackage{float}
\DeclareUnicodeCharacter{2212}{\ensuremath{-}}

\usepackage{booktabs}				
\usepackage{array}					
\usepackage{multirow}				
\usepackage{siunitx}				
\usepackage{flafter}				
\usepackage{comment}				
\usepackage[usenames,dvipsnames]{xcolor}	
\usepackage{flafter}				
\usepackage{lipsum} 				
\usepackage{setspace}               

\usepackage{multicol}

\usepackage{amsfonts}
\usepackage{soul}
\usepackage{url}
\usepackage{epstopdf}
\usepackage{rotating}
\usepackage{placeins}
\usepackage{pifont}
\usepackage{threeparttable}
\usepackage{makecell}
\usepackage{graphicx}
\usepackage{caption}
\usepackage{subcaption}
\usepackage{overpic}
\usepackage{hyperref}	
\usepackage{cleveref}	
\newcommand{\reportingnote}{\protect\textit{Reporting convention:} residuals are summed in absolute value over all constraints of a given type per instance, then averaged over the test set; values are in MVA ($S_{\text{base}}=100$ ~MVA), not p.u. IPOPT's nonzero power-balance entries reflect termination at its default constraint violation tolerance \texttt{constr\_viol\_tol}$=10^{-4}$~p.u. ($10^{-2}$~MVA at $S_{\text{base}}=100$~MVA), rather than exact feasibility.}

\journal{Energy \& AI}

\begin{document}

\begin{frontmatter}



\title{Physics-Informed Graph Neural Networks for Robust AC-Optimal Power Flow}

\author[RRE]{Anna Varbella}
\author[RRE]{Blazhe Gjorgiev}
\author[RRE]{Damien Briens}
\author[SIS]{Giuseppe Alessio D'Inverno}
\author[MIT]{Priya L. Donti}
\author[RRE]{Giovanni Sansavini\corref{cor1}} 

\ead{sansavig@ethz.ch}

\affiliation[RRE]{organization={Reliability and Risk Engineering Laboratory, Institute of Energy and Process Engineering, Department of Mechanical and Process Engineering, ETH Zurich},
            city={Zurich},
           country={Switzerland}}

\affiliation[SIS]{organization={MathLab, International School for Advanced Studies, Trieste}, country={Italy}}
\affiliation[MIT]{organization={Laboratory for Information \& Decision Systems and Department of Electrical Engineering \& Computer Science, Massachusetts Institute of Technology},,
            city={Zurich},
           country={USA}}
\cortext[cor1]{Corresponding author.}

\begin{abstract}
We present PINCO, an unsupervised learning framework that integrates Graph Neural Networks with physics-informed neural networks for AC optimal power flow (AC-OPF) solutions. Unlike state-of-the-art unsupervised methods that require prescreened datasets containing only feasible instances, our approach operates on unfiltered data, including ill-conditioned cases. The framework addresses two critical gaps in the literature: (1) robustness to topology changes up to N-2 contingencies, and (2) detecting optimal power flow instances that are infeasible without relying on traditional solvers for data filtering. PINCO embeds physical laws into the learning process via augmented Lagrangian multipliers. In addition, it introduces a clustering branch with learnable centroids that automatically separate feasible from infeasible solutions based on constraint-violation patterns. We evaluate the framework across systems of varying complexity, including the IEEE 30-bus, IEEE 57-bus, and Swiss transmission networks, demonstrating scalability and robustness under diverse loading conditions and topology variations. Benchmarking against DeepOPF-FT and the IPOPT solver shows that PINCO achieves comparable constraint satisfaction while delivering two to three orders of magnitude computational speedup compared to IPOPT and lower operational costs across all configurations. 
\end{abstract}


\end{frontmatter}


\section{Introduction}
\label{intro}

Power systems must continuously balance electricity generation and demand while keeping voltages and frequency within operational limits. This real-time balance is described by the nonlinear sinusoidal power flow (PF) equations, which capture the steady-state behavior of the electric grid. Beyond merely satisfying physical laws, however, system operators must also allocate the generation of power in a cost-effective way, leading to the optimal power flow (OPF) problem. OPF seeks the most economical power generation dispatch subject to the power flow equations and operational constraints. When the full alternating-current formulation is retained without linearization or other simplifications, the resulting problem is known as the alternating-current optimal power flow (AC-OPF). AC-OPF is a cornerstone of power system operation. This problem is solved in real time, often every five minutes~\cite{Nair2024ACOPF}, as well as in long-term planning studies.

AC-OPF is inherently nonconvex and NP-hard~\cite{opftheory}. Traditionally, different approaches are used to solve OPF. The simplest method, DC-OPF, uses a linear form of the power flow equations. This linear approximation  leads to sub-optimal, infeasible solutions to the complete AC-OPF~\cite{kile2014, DCvsAC}. Alternatively, convex relaxation techniques have been extensively examined~\cite{convexACOPF,opf_relaxation_1,bose2025presolving}. Currently, non-linear programming methods, such as the interior-point method, are commonly used for solving  OPF~\cite{zimmerman2016matpower}. Nevertheless, these methods still face significant computational burdens and scalability challenges when applied to large-scale systems~\cite{Bienstock2022}. 


The limitations of traditional methods have driven researchers to explore alternative solutions using machine learning techniques. For a comprehensive overview of the broader landscape, we refer readers to~\cite{khaloie2025review, MLOPFWiki2025, wolgast2024learning}. Significant effort has been directed at generalizing AC-OPF solvers across varying network topologies. Supervised approaches, including DeepOPF~\cite{deepopf} and its topology-flexible extension DeepOPF-FT~\cite{deepopf-ft}, encode network configurations to handle multiple topologies within a single model, achieving strong accuracy and inference speed. Graph neural networks (GNNs) have further 
advanced this direction by naturally representing power system topology \cite{canos, topaware}, and recent work has extended GNN-based solvers to N-1 contingency scenarios~\cite{piGnnpert}. However, these supervised methods share a structural limitation: they rely entirely on labeled data generated by conventional solvers, binding their generalization capacity to the distribution 
of pre-solved examples. Unsupervised GNN approaches~\cite{owerko2022unsupervisedoptimalpowerflow} reduce label dependence by embedding physics constraints directly into training, and in principle can accommodate topology perturbations, yet this capability has not been rigorously validated.

A parallel limitation concerns feasibility. Unsupervised methods avoid labeled data by incorporating optimality conditions, such as KKT conditions or Lagrangian duality~\cite{selfsupervised, lagrangian-duality, lagrangian-approach}, directly into the loss function. Methods such as DC3~\cite{dc3} and FSNet~\cite{nguyen2025fsnet} go further, enforcing constraint satisfaction by construction during inference. Despite these advances, a subtle but critical contradiction persists: virtually all existing methods, supervised and unsupervised alike, are trained exclusively on prescreened datasets from which ill-conditioned or infeasible operating points have been removed using conventional solvers. For unsupervised methods, this directly undermines their physics-based framing. To the best of our knowledge, no method for AC-OPF has been designed to detect infeasible problem instances jointly with solution prediction; the closest related work~\cite{mohammadian2025restoring} addresses feasibility detection and restoration in the linearised DC-OPF setting via counterfactual explanations, and explicitly identifies AC-OPF generalisation as future work.

Across unsupervised methods for the AC-OPF, therefore, two critical gaps still exist in the literature: (1) no method 
robustly generalizes across different topology changes without dependence on solver-generated labels for each configuration, which limits applicability in real-world systems where configurations frequently change due to contingencies, maintenance, or upgrades; and (2) the absence of mechanisms to detect infeasibility. Indeed, existing unsupervised methods train exclusively on prescreened, well-conditioned datasets where ill-conditioned or infeasible cases have been filtered out using traditional solvers.

Our work addresses these challenges by leveraging physics-informed neural networks with hard constraints (hPINN) \cite{lu2021physics} and Graph Neural Networks (GNNs) to solve the AC-OPF. We call this combined architecture PINCO. The power system physics is retained via embedding physical laws directly into the learning process through augmented Lagrangian multipliers but without strict feasibility enforcement of the power flow equations. Solvable and unsolvable scenarios can be identified via introducing a clustering branch that learns to separate graph-level embeddings into two clusters using learnable centroids. Topological changes can be handled via training on a diverse set of grid configurations, achieving robustness to up to two simultaneous component failures (N−2). We evaluate our approach on two benchmark test systems widely used in prior work, IEEE 30-bus~\cite{ieee30bus} and IEEE 57-bus~\cite{ieee57bus} cases, and on a real-world case study: the Swiss transmission grid~\cite{swissgrid}. We benchmark our results against the Deep OPF-FT model \cite{deepopf-ft} on the IEEE 57-bus system, selecting this baseline as it similarly handles topology variations up to N-2 contingencies.

In this work, we contribute to the ongoing research on physics-informed neural networks for AC-OPF problems as follows:
\begin{itemize}
    \item We combine GNNs with constraint awareness within an unsupervised training methodology that does not require pre-solved AC-OPF instances, eliminating the need for prescreened datasets and enabling operation on unfiltered input data, including ill-conditioned cases.
    
    \item We introduce a learnable clustering branch integrated into the neural network architecture that automatically distinguishes between feasible and infeasible solutions during training. By learning centroids via physics-based contrastive learning, the model separates graph-level embeddings based on constraint-violation patterns, enabling instance-level infeasibility detection.
    
    \item We develop a GNN architecture capable to generalize across varied loading conditions and up to N-2 topology variations, demonstrating robustness to topology changes commonly encountered in real-world operations.
    
    \item Our approach produces models with two to three orders of magnitude computational speedup compared to traditional interior-point solvers. At the same time, it maintains solution quality across benchmark systems and a real-world transmission grid case study.
\end{itemize}


The paper is structured as follows: Section~\ref{sec:methodology} outlines the methodology and problem formulation;  Section~\ref{sec:datasetpinco} introduces the experimental setting and the case studies;  Section~\ref{sec:results} presents the results;  Section~\ref{sec:limitations} presents the limitations of this study;  Section~\ref{sec:conclusions} concludes with final remarks and discussion.


\section{Methodology}
\label{sec:methodology}
This section presents the AC-OPF formulation (Section~\ref{sec:optimalpowerflow}), our physics-informed neural network for constraint optimization (PINCO) approach to solve AC-OPF (Section~\ref{sec:pinnf}), and constraint violation metrics we utilize in PINCO (Section~\ref{sec:pincoarchitecture}). PINCO integrates three key components: physics-informed neural networks (PINNs) that embed physical laws directly into the learning process (Section~\ref{sec:pinn}), Graph Neural Networks (GNNs) that naturally represent power grid topologies and enable generalization across topology variations (Section~\ref{sec:gnn}), and a novel learnable clustering branch that provides instance-level feasibility detection (Section~\ref{sec:pincoarchitecture}).

\subsection{Problem Formulation: AC Optimal Power Flow}
\label{sec:optimalpowerflow}
OPF is a fundamental optimization problem that determines cost-efficient generator dispatch while meeting system constraints in power systems~\cite{opftheory}.
A power system involves a network of electrical buses (i.e., connection points between power lines, where load or generators can be located), denoted by \( N \), interconnected by \( E \) branches (i.e., power lines or transformers). Each bus may host one or more generators, which inject power into the system, and loads, which consume it. The objective of the OPF problem is to minimize the total generation cost while adhering to the system's physical constraints. In this work, we address the full alternating current (AC) version of this model, which better reflects real-world grid conditions. Specifically, we consider a set of generator buses \( N_g \) and load buses \( N_d \). 
The AC-OPF problem can be formulated as follows, where \( C_r \) represents the cost associated with operating generator \( r \):
\begin{equation}
\begin{aligned}
    \min_{P_{g,i}, Q_{g,i}, V_i, \theta_i} \quad & \sum_{r\in N_g} C_r (P_{g,r}) \quad  \forall r \in N_g \\
    \text{s.t.} \quad & h_i(P_{g,i}, Q_{g,i}, V_i, \theta_i, P_{d}, Q_{d}) = 0 \quad \forall h_j \in H \\
    & g_j(P_{g,i}, Q_{g,i}, V_i, \theta_i, P_{d}, Q_{d}) \leq 0 \quad \forall g_l \in G.
\end{aligned}
\end{equation}

The equality constraints \( H \) represent the nodal balance equations:
\begin{equation}
\begin{aligned}
H = \bigcup_{i \in N} &\left\{ P_{g,i} - P_{d,i} - g_{i}^{sh} = \sum_{(ij)\in E} p_{ij} \right\} \\
&\cup \left\{ Q_{g,i} - Q_{d,i} + b_{i}^{sh} = \sum_{(ij)\in E} q_{ij} \right\}.
\end{aligned}
\end{equation}
Here, the active power demand and generation at bus \( i \) are denoted by \( P_{d,i} \) and \( P_{g,i} \) (in MW), while the reactive power demand and generation are \( Q_{d,i} \) and \( Q_{g,i} \) (in MVAr). The active power \( p_{ij} \) and reactive power \( q_{ij} \) flowing between buses \( i \) and \( j \) are governed by the power flow equations:
\begin{align}
    p_{ij} &= g_{ij} (\tau_{ij} V_i^2) - V_i V_j \left( b_{ij} \sin(\theta_{ij}) + g_{ij} \cos(\theta_{ij}) \right) \notag \\
    q_{ij} &= -(g_{ij} + \frac{y^{sh}_{ij}}{2}) (\tau_{ij} V_i^2) - V_i V_j \left( g_{ij} \sin(\theta_{ij}) - b_{ij} \cos(\theta_{ij}) \right).
\end{align}

Here, $V_i$ is the voltage magnitude at bus $i$ (in p.u.) and $\theta_i$ is the voltage phase angle (in rad), with $\theta_{ij} = \theta_i - \theta_j$. The grid parameters include the conductance $g_{ij}$ and susceptance $b_{ij}$ of the transmission lines (in p.u.), shunt admittances $y^{sh}_{ij}$ (in p.u.), and transformer tap ratios $\tau_{ij}$ (dimensionless). Bus shunt elements are characterized by $g_{i}^{sh}$ and $b_{i}^{sh}$, specified following the MATPOWER convention~\cite{matpower} as equivalent MW consumed and MVAr injected at a nominal voltage of $V_i = 1$~p.u., and represent fixed admittances to ground.

The inequality constraints on the generator limits, for active and reactive power at each bus, are defined as:

\begin{align}
G_P &= \bigcup_{r \in N_g} \left\{ P_{G,\min} \leq P_g \leq P_{G,\max} \right\} \notag \\
G_Q &= \bigcup_{r \in N_g} \left\{ Q_{G,\min} \leq Q_g \leq Q_{G,\max} \right\}.
\end{align}
Similarly, the inequality constraints on voltage magnitudes $G_V$ and branch thermal limits $G_S$ are:
\begin{align}
G_V &= \bigcup_{i \in N} \left\{ V_{i,\min} \leq V_i \leq V_{i,\max} \right\} \notag \\
G_S &= \bigcup_{(ij)\in E} \left\{ (p_{ij})^2 + (q_{ij})^2 \leq |S_{\max,ij}| \right\},
\end{align}
where \( |S_{\max,ij}| \) represents the module of the maximum allowable apparent power\footnote{The apparent power $S$ in an AC circuit is the complex quantity $S = VI^*$, where $V$ is the complex voltage phasor and $I^*$ is the complex conjugate of the current phasor. Its magnitude is given by $|S| = \sqrt{P^2 + Q^2}$, where $P$ and $Q$ are the real and reactive power respectively.} transferred between buses \( i \) and \( j \). 
Thus, the total inequality constraints \( G \) are given by the union of the constraints defined above:
\begin{equation}
 G = G_P \cup G_Q \cup G_V \cup G_S.   
\end{equation}

\subsection{Problem solution: Physics-Informed Neural Network for Constraint Optimization (PINCO)}
\label{sec:pinnf}
We formulate AC-OPF as a physics-informed neural network (PINN) problem to enable learning from both feasible and infeasible power system states without requiring pre-solved datasets. PINNs~\cite{raissi2019physics} embed physical constraints directly into the loss function in a way that is compatible with automatic differentiation, allowing the neural network to learn solutions that are penalized for violating governing equations and operational limits. 

\subsubsection{Physics-Informed Neural Networks}
\label{sec:pinn}
Traditional PINNs solve differential equations by minimizing residuals of governing PDEs. However, AC-OPF is fundamentally an optimization problem that requires the simultaneous handling of an objective function (generation cost minimization) and operational constraints (power balance equations, generator limits, voltage limits, thermal limits). To address this problem, we adapt the physics-informed optimization framework proposed in~\cite{lu2021physics}, which extends PINNs to constrained optimization through Augmented Lagrangian methods. Our formulation differs from the original hPINN framework by omitting the quadratic penalty terms. This design choice simplifies the loss weight update schedule by relying solely on gradient-based weighting, in the spirit of methods such as the Neural Tangent Kernel (NTK)~\cite{wang2022and}. At each training iteration $k$, our network minimizes the following loss function to identify the solution parameters $\mathbf{w}_u$:

\begin{equation}
\label{eq:physicsinformedloss}
   \mathcal{L}_r^k(\mathbf{w}_u^k)  =  \mathcal{J}(\mathbf{w}_u^k) + \frac{1}{|H|} \sum \limits_{i=1}^{|H|} \lambda_{h_i}^k |h_i(\mathbf{w}_u^k)| + \frac{1}{|G|} \sum \limits_{j=1}^{|G|} \lambda_{g_j}^k \max(0, g_j(\mathbf{w}_u^k)) 
\end{equation}

\noindent where $\mathcal{J}$ represents the generation cost objective, $h_l$ denotes the $l$-th equality constraint (power balance equations), and $g_j$ denotes the $j$-th inequality constraint (generation, voltage, and thermal limits). The total number of equality constraints is $L = |H| = 2|B|$ (two power balance equations per bus, where $|B|$ is the number of buses), and the total number of inequality constraints is $J = |G| = 4|N_g| + 2|B| + |E|$ (encompassing generator limits, voltage limits, and thermal capacity constraints). The Lagrange multipliers $\lambda_{h_l}^k, \lambda_{g_j}^k$ are updated based on constraint gradient directions~\cite{lu2021physics}. The update mechanism is reported in \ref{apx:al}. This formulation enables the network to learn from instances regardless of their initial feasibility, eliminating the dependency on pre-screened training datasets that burden supervised approaches and some hard-constrained unsupervised approaches.

\subsubsection{Graph Neural Networks integration}
\label{sec:gnn}
To implement this PINN formulation, we employ Graph Neural Networks (GNNs) as the underlying architecture. Power grids are naturally represented as graphs, with nodes corresponding to buses and edges to transmission lines. GNNs are a deep learning architecture that readily adapts to this data structure, allowing information to propagate through the network via message-passing layers. This graph-based representation offers a critical advantage for power system optimization: a single trained model can handle topological variations of the same grid, such as N-1 and N-2 contingency scenarios where lines are removed, without requiring retraining or architectural modifications~\cite{VARBELLA2023, canos}. The GNN processes the grid state, characterized by load conditions $(P_d, Q_d)$ and constant parameters such as generation limits, voltage bounds, thermal limits, and branch admittances. It aims to produce node-level embeddings that encode both local bus characteristics and the global system state. These embeddings are subsequently used by our model architecture as follows.

\begin{figure}[htbp]
    \centering    
    \includegraphics[width=0.45\textwidth]{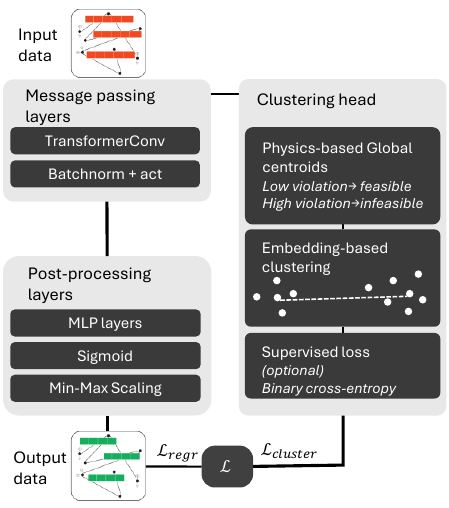}    
    \caption{ The architecture processes power grid data ($P_d$, $Q_d$, Type) using TransformerConv~\cite{TransformerConv} message-passing layers, followed by batch normalization and activation functions. The model employs a dual-branch learning approach: (1) a regression branch with post-processing layers that predict optimal generation ($P_g$, $Q_g$) and electrical parameters ($V$, $\theta$) at each node, and (2) a clustering branch that performs global graph pooling and learns to separate feasible from infeasible instances using learnable centroids. The combined loss function ($\mathcal{L}$) integrates regression loss ($\mathcal{L}_r$) for solution accuracy and clustering loss ($\mathcal{L}_{\text{cluster}}$).}  
    \label{fig:PIGNN}
\end{figure}

\subsubsection{Dual-Branch Graph Neural Network Architecture}
\label{sec:pincoarchitecture}
We design a dual-branch architecture to address a fundamental challenge of unsupervised AC-OPF learning: distinguishing between feasible solutions and infeasible solutions without pre-screening input scenarios.
Figure~\ref{fig:PIGNN} shows a graphical representation of the design with a data processing (top left), a \textbf{solution branch} (bottom left), and a \textbf{clustering branch} (right). The full architecture hyperparameters are in \ref{apx:hyp}. 
The power system data is processed using TransformerConv message-passing layers~\cite{TransformerConv}, followed by batch normalization and activation functions. This setup produces node-level embeddings that capture the system's electrical state and topology.
These embeddings are then processed through the two branches. The regression branch applies post-processing layers to predict optimal generation setpoints $(P_g, Q_g)$ and electrical parameters $(V, \theta)$ at each node. It uses a sigmoid layer to enforce inequality constraints on voltage magnitude and active and reactive power of generators. The branch is trained via the loss defined in Equation~\ref{eq:physicsinformedloss}. 

While the regression branch optimizes power flow solutions, the clustering branch enables the model to identify whether these regression outputs represent feasible or infeasible solutions.
During unsupervised training, the network encounters both feasible instances and infeasible instances that violate operational constraints (e.g., insufficient generation capacity or thermal limits under high loading). To distinguish these cases at inference, we integrate a learnable clustering module that separates instances based on their constraint violation patterns. Unlike post-hoc approaches that apply fixed thresholds to violation metrics after training, our method learns the feasibility boundary jointly with the regression task, allowing the clustering and regression branches to co-adapt throughout the optimization process.

The clustering module operates on graph-level embeddings $\mathbf{z} \in \mathbb{R}^d$ obtained through global additive pooling of node features from the message-passing layers. Two learnable centroids $\mathbf{c}_1, \mathbf{c}_2 \in \mathbb{R}^d$ are initialized randomly and optimized through backpropagation alongside all other network parameters. Both embeddings and centroids are L2-normalized. For each graph embedding $\mathbf{z}$, we compute Euclidean distances to the normalized centroids:
\begin{equation}
d_i = \|\mathbf{z} - \mathbf{c}_i\|_2, \quad i \in \{1, 2\}.
\end{equation}
Soft cluster assignments are obtained through a temperature-scaled softmax:
\begin{equation}
p_i = \frac{\exp(-d_i / \tau)}{\sum_{j=1}^{2} \exp(-d_j / \tau)}
\end{equation}
where $\tau = 0.5$ controls assignment sharpness.

The clustering loss function combines four complementary objectives to learn meaningful feasibility boundaries:
\begin{equation}
\label{eq:clusteringloss}
\mathcal{L}_{\text{cluster}} = \mathcal{L}_{\text{compact}} + \mathcal{L}_{\text{sep}} + \mathcal{L}_{\text{orth}} + \mathcal{L}_{\text{physics}}
\end{equation}
where $\mathcal{L}_{\text{compact}} = \mathbb{E}[\sum_{i} p_i d_i^2]$ enforces compactness by minimizing within-cluster distances, $\mathcal{L}_{\text{sep}} = -\|\mathbf{c}_1 - \mathbf{c}_2\|_2$ maximizes separation between centroids, $\mathcal{L}_{\text{orth}} = \langle \mathbf{c}_1, \mathbf{c}_2 \rangle^2$ encourages orthogonality between cluster representations to prevent degenerate solutions, and $\mathcal{L}_{\text{physics}}$ is a physics-based contrastive loss that groups embeddings with similar constraint violation profiles.

The physics-based contrastive component $\mathcal{L}_{\text{physics}}$ guides clustering using constraint violation information without requiring ground-truth feasibility labels from external solvers. To quantify constraint satisfaction in the AC-OPF problem, we define metrics that serve dual purposes: they provide training signals for our physics-informed clustering branch and enable quantitative evaluation of solution quality against conventional IPOPT solutions.

We measure constraint violations across four categories:
\begin{enumerate}
\item \textbf{Equality constraint violations} ($E_{viol}$), which quantify power balance deviations at each bus, where $S$ represents active ($P$) and reactive ($Q$) power, $S_{ij}$ is the power flow from node $i$ to $j$, $N$ is the set of nodes, and $E$ is the set of edges in the power network:
\vspace{0.3cm}
\begin{equation}\label{eq:eq_loss}
E_{viol} = \sum_{S \in {P,Q}}\sum_{i\in N}\sum_{ij\in E} |S_i^{gen} - S_i^{load} - S_{ij}|.
\end{equation}

\vspace{0.3cm}
\item \textbf{Generator inequality constraint violations} ($G_{viol}$), which compute the violations of generator capacity limits, where $G$ is the set of generators, and $P_i^{min}$, $P_i^{max}$, $Q_i^{min}$, $Q_i^{max}$ represent the generation limits: 
\vspace{0.3cm}
\begin{equation}\label{eq:gen_ineq_loss}
\begin{aligned}
    G_{viol} = \sum_{i \in G} \Big( &\max(0, P_i - P_i^{max}) + \max(0, P_i^{min} - P_i) \\
    &+ \max(0, Q_i - Q_i^{max}) + \max(0, Q_i^{min} - Q_i) \Big).
\end{aligned}
\end{equation}

\vspace{0.3cm}
\item \textbf{Node inequality constraint violations} ($N_{viol}$), which measure voltage magnitude bound violations, where $V_i$ is the voltage magnitude at node $i$, with upper and lower bounds $V_i^{max}$ and $V_i^{min}$:
\vspace{0.3cm}
\begin{equation}\label{eq:node_ineq_loss}
    N_{viol} = \sum_{i \in N} \Big( \max(0, V_i - V_i^{max}) + \max(0, V_i^{min} - V_i) \Big).
\end{equation}

\vspace{0.3cm}
\item \textbf{Flow constraint violations} ($F_{viol}$), which represent transmission line capacity exceedances, where $S_{ij}$ is the apparent power flow on line $ij$ and $S_{ij}^{max}$ is its thermal limit: 
\vspace{0.3cm}
\begin{equation}\label{eq:flow_loss}
    F_{viol} = \sum_{ij \in E} \max(0, |S_{ij}| - S_{ij}^{max}).
\end{equation}

\end{enumerate}

We define the total violation as: 
\begin{equation}
v_{\text{total}} = E_{viol} + G_{viol} + N_{viol} + F_{viol}.
\end{equation}
These total violations $v_{\text{total}}$ are normalized within each training batch, to obtain $v_{\text{norm}}$ through mean normalization.
Pseudo-labels are assigned by comparing each instance's normalized violation to the batch median, effectively separating instances into low-violation and high-violation groups:
\begin{equation}
y^{\text{phys}} = \begin{cases}
1, & \text{if } v_{\text{norm}} < \text{median}(v_{\text{norm}}) \\
0, & \text{otherwise}.
\end{cases}
\end{equation}
To maintain consistency across training batches, we track global centroids for low-violation and high-violation groups using an exponential moving average with momentum $\mu = 0.9$. The contrastive loss then encourages embeddings with similar constraint patterns to cluster together while pushing apart those with dissimilar patterns:
\begin{equation}
\begin{split}
\mathcal{L}_{\text{physics}} = \mathbb{E}_{i,j}\bigg[&(1 - \mathbf{z}_i^T\mathbf{z}_j) \cdot \mathbbm{1}[y_i^{\text{phys}} = y_j^{\text{phys}}] \\
&+ \max\left(0, \mathbf{z}_i^T\mathbf{z}_j - 0.5\right) \cdot \mathbbm{1}[y_i^{\text{phys}} \neq y_j^{\text{phys}}]\bigg].
\end{split}
\end{equation}

The complete training objective combines the physics-informed regression loss (Equation~\ref{eq:physicsinformedloss}) and the clustering loss (Equation~\ref{eq:clusteringloss}):
\begin{equation}
\label{eq:completeloss}
\mathcal{L} = \mathcal{L}_r + \mathcal{L}_{\text{cluster}}.
\end{equation}

\textbf{Semi-supervised extension.} To evaluate the potential benefit of incorporating ground-truth feasibility information, we also develop a variant of the complete training objective, where a small percentage of training instances include feasibility labels obtained from IPOPT. In this mode, an additional term $\mathcal{L}_{\text{supervised}}$ is added to $\mathcal{L}$ in Equation~\ref{eq:completeloss}:
\begin{equation}
\mathcal{L}_{\text{supervised}} = -\mathbb{E}_{k \in \mathcal{S}}\left[y_k \log(p_{k,1}) + (1-y_k)\log(1-p_{k,1})\right]
\end{equation}
where $\mathcal{S}$ denotes the subset of labeled instances, $y_k \in \{0,1\}$ is the ground-truth feasibility label from IPOPT, and $p_{k,1}$ is the predicted probability of belonging to the feasible cluster. This binary cross-entropy loss encourages learned cluster assignments to align with ground truth when labels are available. 

\section{Case studies}
\label{sec:datasetpinco}
In this work, PINCO is applied to the IEEE 30-bus, the IEEE 57-bus, and the Swiss power system. 
The IEEE power systems are widely recognized benchmarks frequently used in power systems research. 
%
The IEEE 30-bus system comprises six generator units, 41 transmission lines, and 4 transformers~\cite{ieee30bus}. The IEEE 57-bus system~\cite{ieee57bus} includes seven generator units and 80 branches (63 transmission lines and 17 transformers). The Swiss transmission grid model represents Switzerland's extra-high-voltage transmission network~\cite{swissgrid}, which serves as a critical hub in the European power system due to its central location. The grid includes 180 buses, 309 transmission lines, and 91 generators.

\subsection{Dataset structure}

We model the electrical grid as a graph with a set of nodes $N$ corresponding to electrical buses and edges $E$ representing branches. The graph is defined as $G =(N, E, \textbf{N},\textbf{A}, \textbf{E})$. We call $ \textbf{N} \in \mathbb{R}^{|N|\times t} $ the node feature matrix, with $|N|$ equal to the number of nodes and  \textit{t} to the number of features per node, and $ \textbf{A} \in \mathbb{R}^{|N|\times |N|} $ is the adjacency matrix. The elements of \textbf{A}, $a_{ij}$, are equal to 1 if there is an edge from node \textit{i} to node \textit{j}, and zero otherwise. Finally, $ \textbf{E} \in \mathbb{R}^{|E|\times s} $ is the edge feature matrix, with $|E|$ equal to the number of edges and \textit{s} to the number of features per edge. We predict $\textbf{Y}$ $ \in \mathbb{R}^{|N|\times f} $, the node output matrix, where a vector of $f$ elements is predicted for each node in $N$. 

\begin{figure}[htpb]
  \centering

  \includegraphics[width=0.45\textwidth]{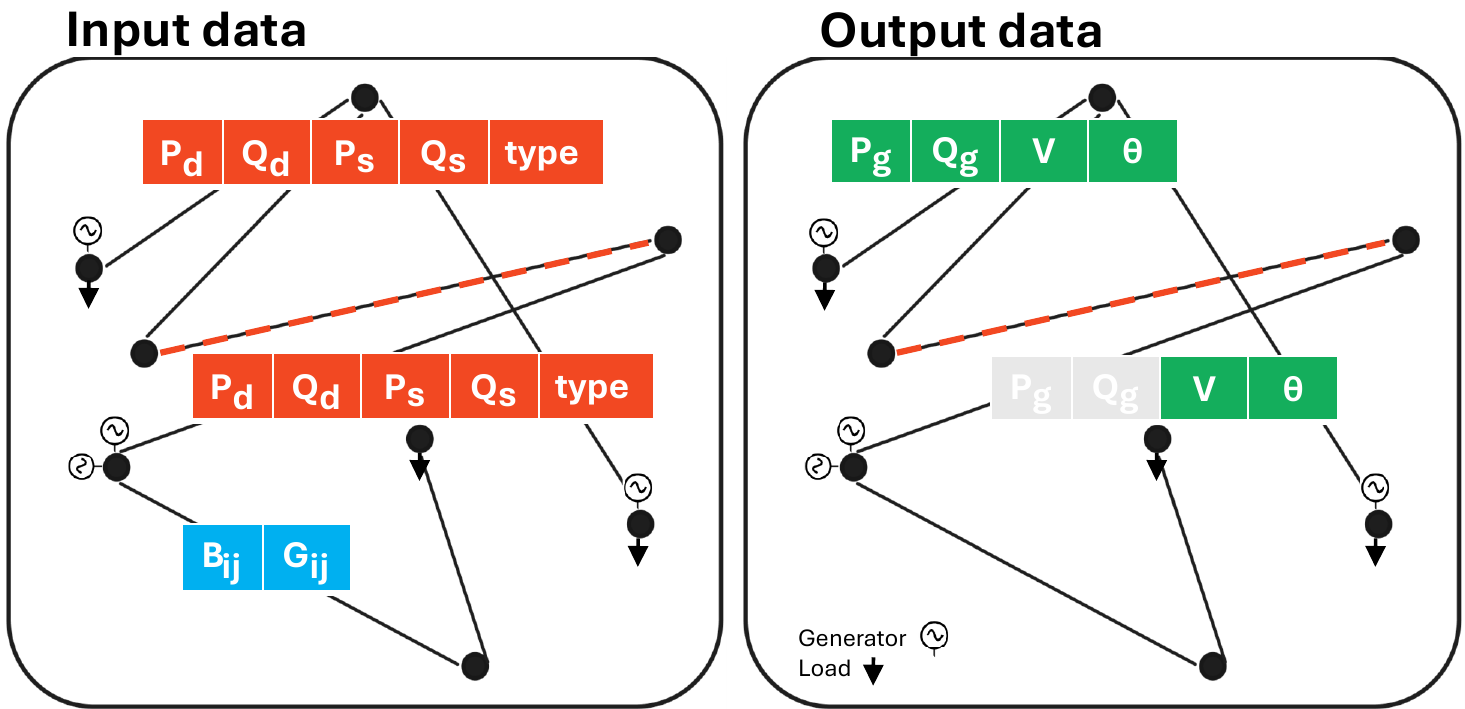}

  \caption{Power grid model as graph. The input features assigned to each node are shown in red, and the predicted quantities at the node level are shown in green. The red dotted line symbolizes a failed branch.}
    \label{fig:datasetpinco}
\end{figure}

Figure \ref{fig:datasetpinco} illustrates the structure of a power grid with its features. The input node features are depicted in red, while the output node-level predictions are in green. The input features include active and reactive power demand ($P_{d}$, $Q_{d}$), active and reactive shunts ($P_s$, $Q_s$), and node type. In total, there are three possible node types, encoded as categorical variables: (1) load bus without a generator, (2) bus with generators, (3) bus with neither load nor generator. The output predictions vary depending on the node type. For example, if a variable is known (e.g., buses without generators do not require predictions of generated power), we do not include it during training. Specifically, $P_g$ and $Q_g$ are absent (grey) for load buses, as these nodes have no generators and therefore no generated power to predict. The predicted output quantities include active power generation ($P_{g}$), reactive power generation ($Q_{g}$), voltage magnitude ($V$), and voltage angle ($\theta$). Notably, the training dataset is completely unlabeled. Figure~\ref{fig:datasetpinco} visualizes the neural network prediction, not a particular ground-truth solution. Finally, the edge features are branch conductance $G_{ij}$ and branch susceptance $B_{ij}$. We consider topology variation up to N-2 branch failures. Thus, the failed branches are removed from both the edge feature matrix and the adjacency matrix, which are updated accordingly.

\begin{table}[h]
\centering
\caption{Summary of grid dataset properties across three test systems, ordered by increasing size. The table shows topological characteristics, convergence rates for optimal power flow solutions, normalization parameters, and dataset partitioning. Convergence rates indicate the percentage of scenarios where IPOPT produces feasible solutions.}
\footnotesize
\label{tab:grid-properties}
\resizebox{0.45\textwidth}{!}{%
\begin{tabular}{l|ccc}
\hline
\textbf{Property} & \textbf{IEEE30} & \textbf{IEEE57} & \textbf{Swissgrid} \\
\hline
Number of Nodes & 30 & 57 & 180 \\
Number of Generators & 6 & 7 & 91 \\
Total Topology Variations & 39 & 74 & 166 \\
\hline
\multicolumn{4}{c}{\textbf{IPOPT Convergence Rates (\%)}} \\
\hline
Test Set Convergence Rate & 71.28 & 51.17 & 97.55 \\
Validation Set Convergence Rate & 67.52 & 51.44 & 97.63 \\
\hline
\multicolumn{4}{c}{\textbf{Normalization Values}} \\
\hline
P norm & 80.0 & 575.88 & 1539.0 \\
Q norm & 150.0 & 200.0 & 720.0 \\
V norm & 1.1 & 1.06 & 1.1 \\
Theta norm & 360 & 360 & 360 \\
Base MVA & 100 & 100 & 100 \\
\hline
\multicolumn{4}{c}{\textbf{Dataset Properties}} \\
\hline
Training Samples & 2730 & 5180 & 11620 \\
Validation Samples & 585 & 1110 & 2490 \\
Test Samples & 585 & 1110 & 2490\\
\hline
\end{tabular}%
}
\end{table}

Our experimental evaluation utilizes datasets constructed from three power systems of varying complexity, summarized in Table~\ref{tab:grid-properties}: the IEEE30, IEEE57, and large-scale Swiss power system networks. For a given reference loading condition, the active and reactive power demands are sampled from uniform distributions centered at 90\% and 110\% of their respective reference values. When simulating contingencies, we employ a strategic sampling approach that varies with grid size: smaller grids (IEEE30 and IEEE57) allow for complete N-1 analysis and 20\% coverage of N-2 scenarios, while the large-scale Swiss power system uses 30\% of N-1 conditions and 0.3\% of N-2 contingencies to maintain computational tractability. This approach generates 100 loading conditions combined with multiple topology variations, resulting in both feasible and infeasible scenarios.

From this primary dataset, we construct two experimental configurations to evaluate different variants of the PINCO model:

\textbf{Configuration 1 (Filtered)} -- A filtered dataset containing only feasible instances serves as a controlled environment to evaluate PINCO in supervised mode (PINCO-F-Sup) and physics-informed unsupervised mode (PINCO-F-Phys), both without clustering mechanisms.

\textbf{Configuration 2 (Unfiltered)} -- An unfiltered dataset including all instances regardless of feasibility enables evaluation of PINCO with clustering-based feasibility detection. This configuration is used to train the full PINCO model in physics-informed unsupervised mode (PINCO-U-Phys) and in semi-supervised mode, where 25\% of instances are labeled with binary feasibility indicators to train the clustering head (PINCO-U-Semi25).

For both configurations, the training, validation, and test datasets were created from a common set of generated inputs, which were split into 70\%, 15\%, and 15\%, respectively. IPOPT solutions are obtained for validation and test sets where convergence is achieved. The training set includes only IPOPT solutions to test supervised ablation of the PINCO method, but no IPOPT-convergence-based pre-filtering is applied in any OPF scenario.

\subsection{Evaluation metrics}
\label{sec:evalua}

For evaluation purposes, we define two additional metrics. First, the \textbf{cost optimality gap}, measuring the percentage difference between our method's cost and the IPOPT solution cost:
\begin{equation}\label{eq:cost_gap}
\text{Gap}_{\%} = \frac{C_{\text{PINCO}} - C_{\text{IPOPT}}}{C_{\text{IPOPT}}} \times 100\%
\end{equation}
where $C_{\text{PINCO}}$ and $C_{\text{IPOPT}}$ represent the objective function values (generation costs) from PINCO and IPOPT, respectively.

PINCO handles both feasible and infeasible instances without preprocessing. Therefore, we also define the \textbf{agreement metric}, which quantifies how well PINCO aligns with IPOPT in classifying solution feasibility:
\begin{equation}\label{eq:agreement}
\text{Agreement} = \frac{N_{\text{ff}} + N_{\text{ii}}}{N_{\text{total}}} \times 100\%
\end{equation}
where $N_{\text{ff}}$ and $N_{\text{ii}}$ are the number of instances classified as feasible and infeasible by both methods, respectively, and $N_{\text{total}}$ is the total number of test instances. PINCO's classification is learned via the clustering branch, while IPOPT solutions are classified as feasible based on the solver's convergence criteria, i.e., a tolerance threshold on constraint violations or maximum iteration limit.

\section{Results}
\label{sec:results}
In this section, we evaluate the performance of the PINCO method across the test datasets of IEEE benchmarks and Swiss power systems. All model experiments are conducted on the GPU nodes of the ETH Euler clusters~\cite{eulerwiki} using NVIDIA RTX 3090 nodes. The results are compared to the IPOPT~\cite{IPM} solver solutions obtained on a Windows desktop with an AMD Ryzen Threadripper 3960X 24-Core Processor.
\newcounter{repnote}
\setcounter{repnote}{\value{footnote}}
\begin{table}[h!]
\centering
\caption{Performance comparison of \mbox{PINCO-F-Phys}, \mbox{DeepOPF-FT}, and IPOPT on the IEEE 57-bus test case. Inference times per sample. IPOPT: 8 parallel workers; PINCO and \mbox{DeepOPF-FT}: batch size 128 on GPU. Residuals are summed in absolute value over constraints per instance, averaged over the test set, and reported in MVA ($S_{\text{base}}=100$~MVA); see footnote~\protect\footnotemark[\value{repnote}] for full details. Results are reported as mean~$\pm$~std over three random seeds.}
\label{tab:performance_comparison}
\setlength{\tabcolsep}{3pt}
\resizebox{\columnwidth}{!}{%
\footnotesize
\begin{tabular}{lcc|c}
\hline
Metric & PINCO-F-Phys & DeepOPF-FT & IPOPT \\
\hline
Power Bal.\ (MVA)$\downarrow$  & $\mathbf{1.05{\cdot}10^{0} \pm 2.59{\cdot}10^{-1}}$ & $4.0{\cdot}10^{0} \pm 6.0{\cdot}10^{-1}$ & $6.5{\cdot}10^{-1}$ \\
Gen Ineq (MVA)$\downarrow$     & $0$ & $2.5{\cdot}10^{-2} \pm 1.5{\cdot}10^{-3}$ & $0$ \\
Node Ineq (p.u.)$\downarrow$   & $0$ & $0$ & $0$ \\
Thermal (MVA)$\downarrow$      & $0$ & $0$ & $0$ \\
Cost (\$)$\downarrow$          & $\mathbf{4.05{\cdot}10^{4} \pm 5.34{\cdot}10^{2}}$ & $4.15{\cdot}10^{4} \pm 2.36{\cdot}10^{2}$ & $4.13{\cdot}10^{4}$ \\
Time/sample (ms)$\downarrow$   & $\mathbf{1.5{\cdot}10^{-1}}$ & $1.779{\cdot}10^{1}$ & $6.44{\cdot}10^{0}$ \\
\hline
\end{tabular}%
}
\footnotetext{\reportingnote}
\end{table}

\footnotetext{\reportingnote}
\subsection{Comparison to DeepOPF-FT on IEEE57}

\setcounter{repnote}{\value{footnote}}
\begin{table*}[t!]
\centering
\caption{Performance comparison of PINCO variants and IPOPT across test sets of IEEE 30-bus, IEEE 57-bus, and Swiss power grid test cases. IPOPT serves as a benchmark. The PINCO models represent 
ablations of the full physics-informed architecture \mbox{PINCO-U-Phys}: \mbox{PINCO-F-Phys} removes the feasibility-detection branch (single-branch, physics loss); \mbox{PINCO-F-Sup} further replaces the physics loss with supervised training; and \mbox{PINCO-U-Semi25} 
restores the full dual-branch architecture but uses semi-supervised training with only 25\% binary labels on feasibility. Bold entries indicate the best-performing neural model per metric and test case. Residuals are summed in absolute value over constraints per instance, averaged over the test set, and reported in MVA ($S_{\text{base}}=100$~MVA); see footnote~\protect\footnotemark[\value{repnote}] for full details. Seed robustness analysis for selected configurations is reported in ~\ref{app:seeds}.}
\label{tab:combined_comparison}
\footnotesize
\begin{tabular}{llcccccc}
\hline
Case & Model & Power Bal. (MVA)$\downarrow$ & Thermal (MVA)$\downarrow$ & Cost (\$) $\downarrow$ & Gap\%$\downarrow$ & Agreement\% \\
\hline\hline
\multirow{5}{*}{IEEE30} 
& IPOPT & $2.89\cdot10^{-1}$ & $5.02\cdot10^{-3}$ & $5.81\cdot10^{2}$ & -- & -- \\
\cline{2-7}
& PINCO-F-Sup & $1.79$ & $2.76\cdot10^{-1}$ & $\mathbf{5.58\cdot10^{2}}$ & $\mathbf{-3.97}$ & -- \\
& PINCO-F-Phys & $6.1\cdot10^{-1}$ & $1.73\cdot10^{-2}$ & $5.80\cdot10^{2}$ & $-0.18$ & -- \\
& PINCO-U-Semi25 & $6.3\cdot10^{-1}$ & $\mathbf{0}$ & $5.73\cdot10^{2}$ & $-1.34$ & $\mathbf{77.61}$ \\
& PINCO-U-Phys & $\mathbf{4.8\cdot10^{-1}}$ & $3.44\cdot10^{-3}$ & $5.67\cdot10^{2}$ & $-2.47$ & $76.07$ \\
\hline\hline
\multirow{5}{*}{IEEE57} 
& IPOPT & $6.51\cdot10^{-1}$ & $0$ & $4.18\cdot10^{4}$ & -- & -- \\
\cline{2-7}
& PINCO-F-Sup & $2.45$ & $\mathbf{0}$ & $4.13\cdot10^{4}$ & $-1.38$ & -- \\
& PINCO-F-Phys & $1.23$ & $\mathbf{0}$ & $\mathbf{4.02\cdot10^{4}}$ & $\mathbf{-3.88}$ & -- \\
& PINCO-U-Semi25 & $\mathbf{9.0\cdot10^{-1}}$ & $\mathbf{0}$ & $4.06\cdot10^{4}$ & $-3.03$ & $\mathbf{68.92}$ \\
& PINCO-U-Phys & $1.06$ & $\mathbf{0}$ & $4.05\cdot10^{4}$ & $-3.22$ & $66.31$ \\
\hline\hline
\multirow{5}{*}{Swiss} 
& IPOPT & $1.84$ & $3.43\cdot10^{-2}$ & $7.03\cdot10^{4}$ & -- & -- \\
\cline{2-7}
& PINCO-F-Sup & $1.69\cdot10^{1}$ & $\mathbf{0}$ & $7.13\cdot10^{4}$ & $1.53$ & -- \\
& PINCO-F-Phys & $2.93$ & $\mathbf{0}$ & $6.99\cdot10^{4}$ & $-0.48$ & -- \\
& PINCO-U-Semi25 & $\mathbf{2.44}$ & $\mathbf{0}$ & $6.99\cdot10^{4}$ & $-0.46$ & $\mathbf{81.77}$ \\
& PINCO-U-Phys & $2.51$ & $\mathbf{0}$ & $\mathbf{6.90\cdot10^{4}}$ & $\mathbf{-1.82}$ & $68.96$ \\
\hline
\end{tabular}
\end{table*}

\setcounter{repnote}{\value{footnote}}
\begin{figure*}[h!]
    \centering
    
    \begin{subfigure}[b]{0.32\textwidth}
        \centering
        \includegraphics[width=\textwidth]{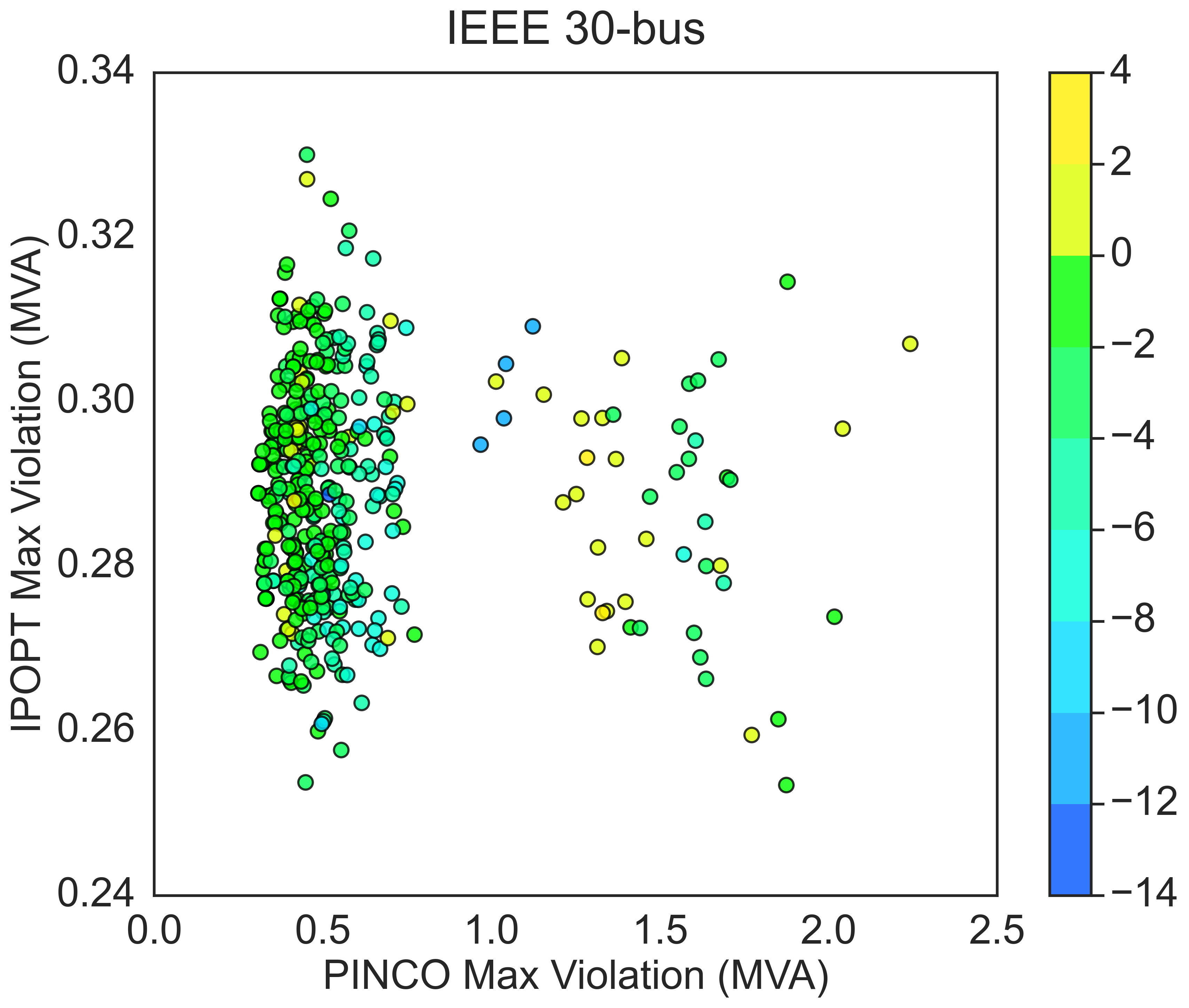}
        \caption{}
        \label{fig:cost_gap_ieee30}
    \end{subfigure}
    \hfill
    \begin{subfigure}[b]{0.32\textwidth}
        \centering
        \includegraphics[width=\textwidth]{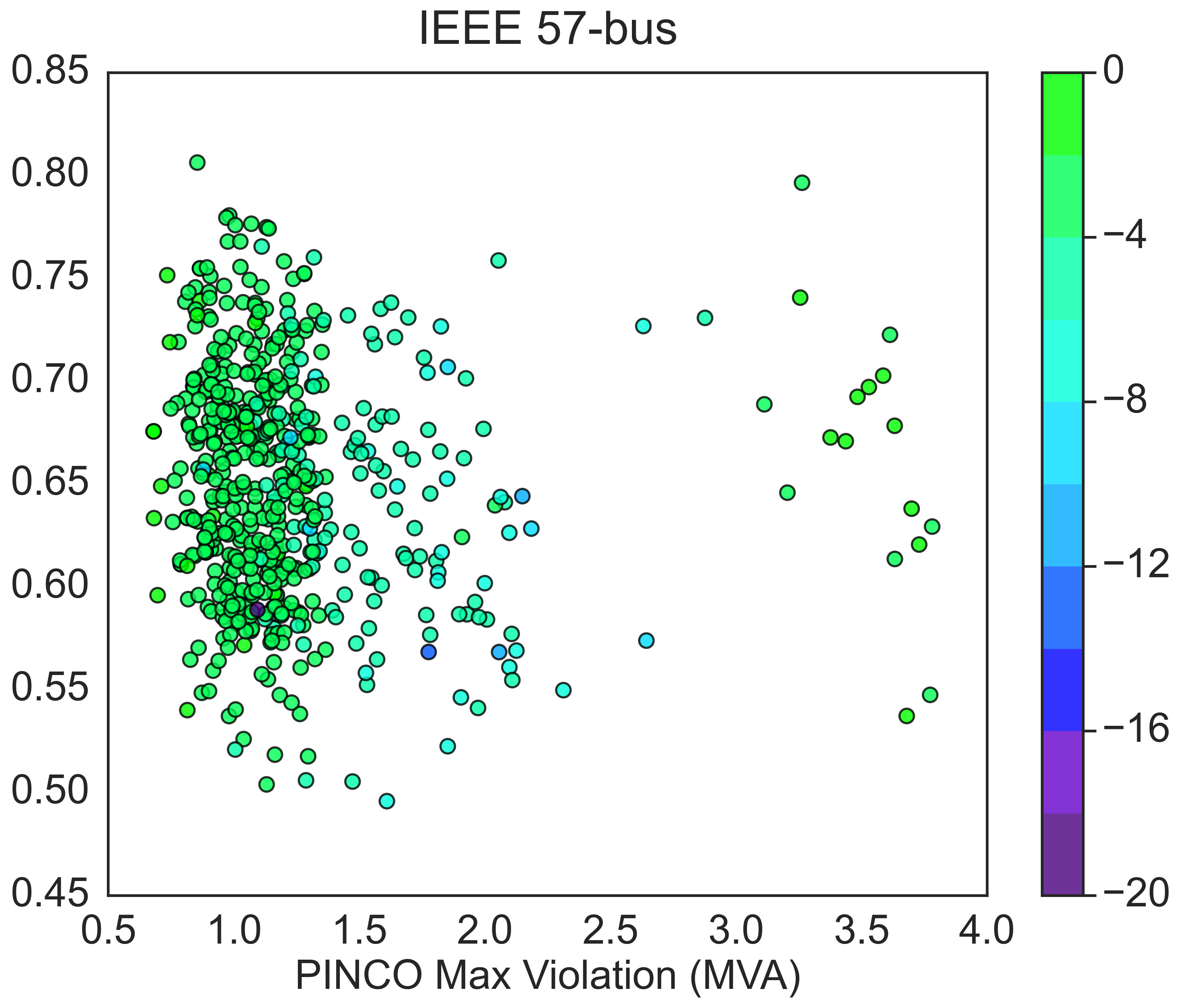}
        \caption{}
        \label{fig:cost_gap_ieee57}
    \end{subfigure}
    \hfill
    \begin{subfigure}[b]{0.32\textwidth}
        \centering
        \includegraphics[width=\textwidth]{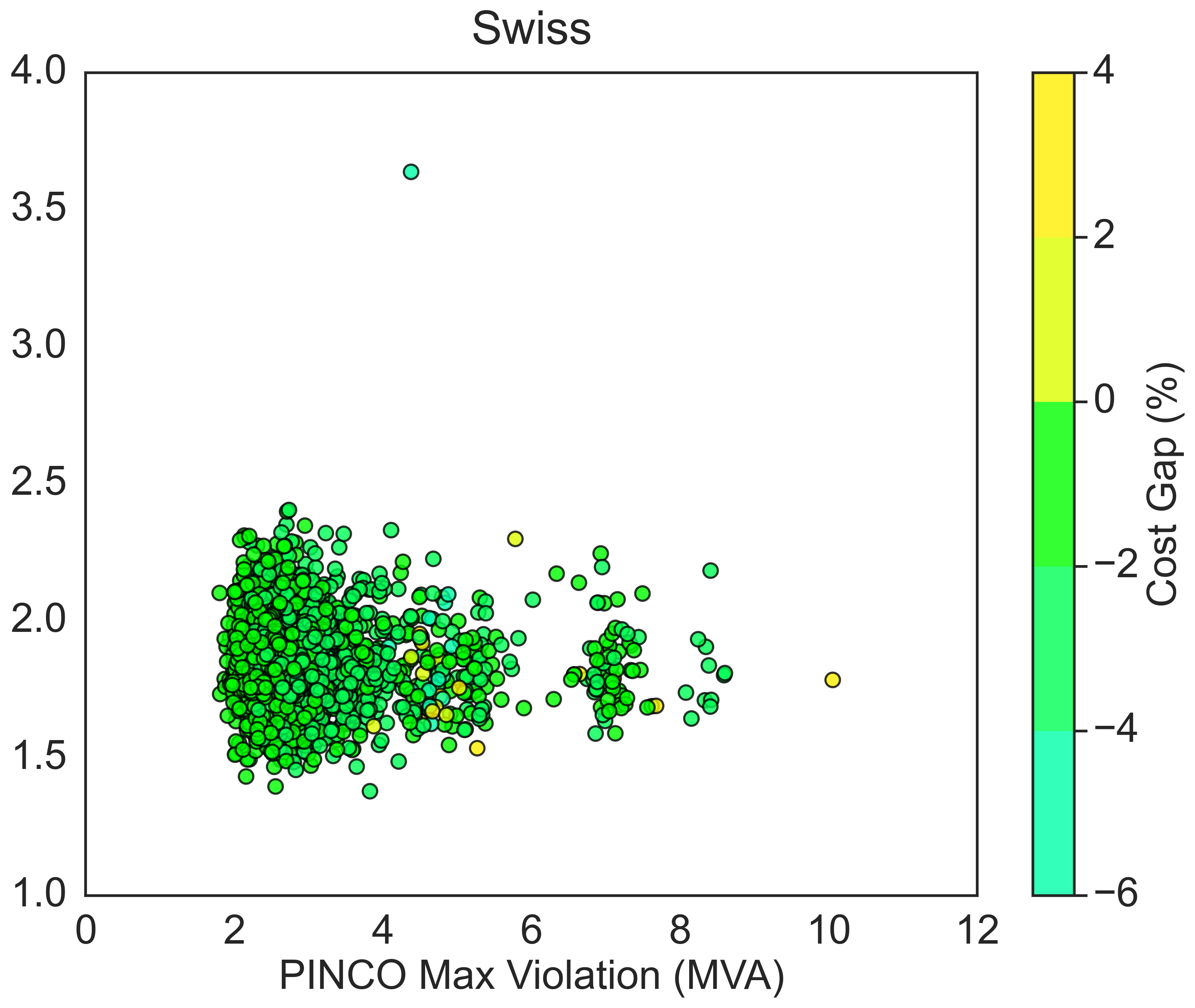}
        \caption{}
        \label{fig:cost_gap_swiss}
    \end{subfigure}
    
    \caption{Cost gap (\%) versus power balance violation (MVA, summed over buses
per instance, $S_{\text{base}}=100$~MVA) across test set samples for the 
IEEE 30-bus (a), IEEE 57-bus (b), and Swiss grid (c) test cases. Each point represents a single test sample, mutually feasible in IPOPT and PINCO-U-Phys. The Cost gap colorbars are consistent across panels (a)-(c). This visualization highlights the trade-off between constraint feasibility and cost optimality at the sample level, 
going beyond aggregated metrics to reveal the spread and concentration of violations across operating conditions.}
    \label{fig:cost_gap_all_grids}
\end{figure*}

We compare our method to DeepOPF-FT, which is a supervised learning approach that considers topology variations up to N-2 contingencies. The hyperparameter values for both methods are in \ref{apx:hyp}. DeepOPF-FT is trained exclusively on feasible instances. Therefore, we compare its performance against PINCO-F-Physics (our unsupervised method trained only on filtered feasible instances) and the IPOPT solver. DeepOPF-FT employs a supervised learning approach with a predict-and-reconstruct mechanism. It first trains a multi-layer feed-forward DNN to predict bus voltages (magnitude and angle) by mapping load and admittance to the AC-OPF solution using pre-solved ground-truth instances. The remaining variables (active and reactive power generation, Pg and Qg) are then reconstructed through using the power flow equations. After the reconstruction mechanism, DeepOPF-FT still requires post-processing corrections to project solutions onto the feasible region and ensure full constraint satisfaction. The results in Table~\ref{tab:performance_comparison} demonstrate a significant computational advantage of our approach: \mbox{PINCO-F-Phys} achieves inference times of 0.15~ms, making it $115\times$ faster than \mbox{DeepOPF-FT}. We note that \mbox{DeepOPF-FT} employs a post-processing correction step that enforces feasibility via an embedded solver, which provides stronger feasibility guarantees than loss-based constraint penalisation; the speed difference partly reflects this architectural trade-off rather than a direct like-for-like comparison. Indeed, \mbox{PINCO-F-Phys} incorporates physics-informed constraints during training and enforces bound constraints by construction, eliminating post-processing while achieving competitive solution quality on this test case. We further note that, as the dataset in the original \mbox{DeepOPF-FT} repository was not directly available, we re-implemented the method following the published description on our dataset; results may therefore differ from those reported in the original work. All residuals in Table~\ref{tab:performance_comparison} are reported in MVA (not p.u.), summed over constraints per instance and averaged over the test set; all reported cases are feasible. Under this convention, \mbox{PINCO-F-Phys} exhibits lower power balance violations and lower generation cost compared to our \mbox{DeepOPF-FT} re-implementation. We omit the analysis of IPOPT computational time as it is further analyzed in Section~\ref{sec:time}.

\subsection{Numerical Results and Model Comparison}
\begin{figure*}[h!]
    \centering
        \includegraphics[width=0.9\textwidth]{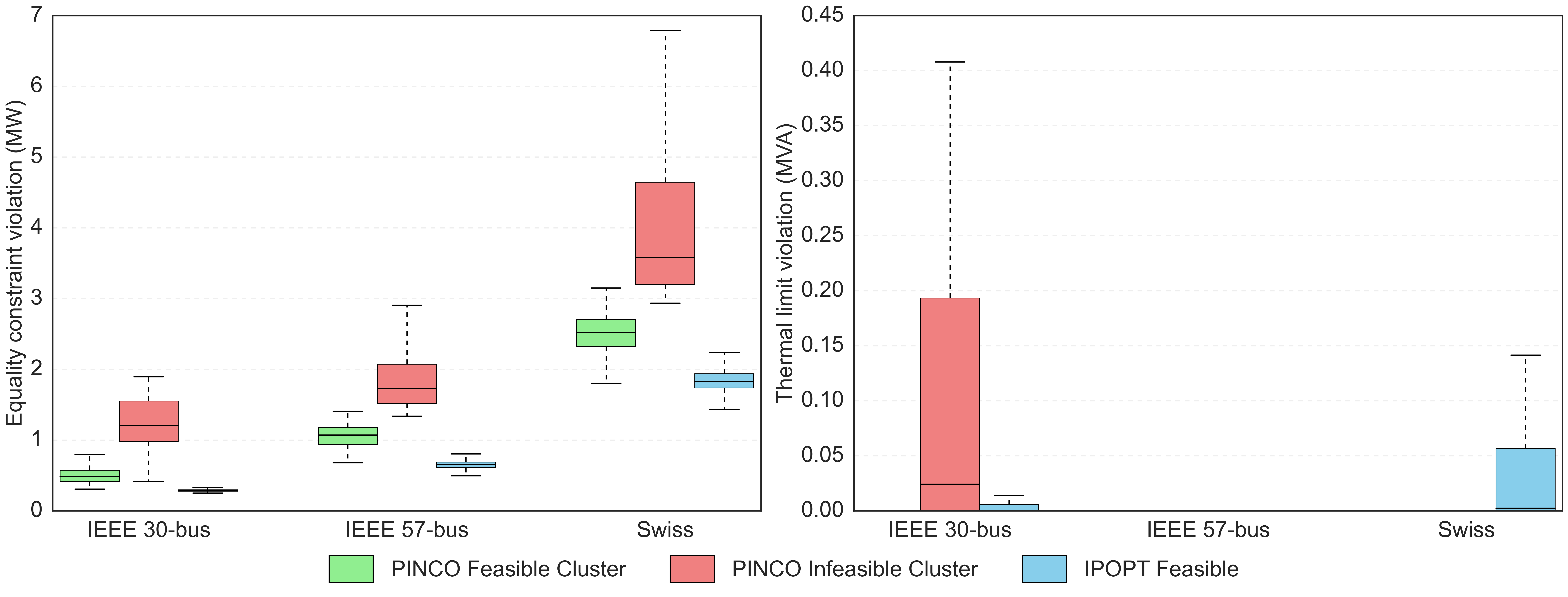}
    \caption{Constraint violation distributions shown as box plots across \mbox{PINCO}-classified feasible and infeasible clusters alongside IPOPT feasible solutions, for the test datasets of the IEEE 30-bus, IEEE 57-bus, and Swiss grid test cases. \mbox{PINCO} partitions test samples into a feasible cluster (IEEE30: 457, IEEE57: 712, Swiss: 1752 samples) and an infeasible cluster (IEEE30: 128, IEEE57: 398, Swiss: 738 samples) via its dual-branch architecture. IPOPT solutions are shown only for samples that converged to a feasible solution (IEEE30: 417, IEEE57: 568, Swiss: 2429 samples). Each box plot displays the median and interquartile range of the equality constraint (eq.~\ref{eq:eq_loss}) and thermal limit violation magnitudes (eq.~\ref{eq:flow_loss}). The figure illustrates the internal consistency of \mbox{PINCO}'s dual-branch architecture: the infeasible cluster exhibits systematically higher predicted violations than the feasible cluster, confirming that the solution prediction branch and the feasibility classification branch produce coherent outputs. This is distinct from ground-truth feasibility detection accuracy, which is quantified separately via the Agreement\% metric in Table~\ref{tab:combined_comparison}.}
    \label{fig:violation_comparison}
\end{figure*}
Table~\ref{tab:combined_comparison} compares PINCO variants against the IPOPT benchmark. All PINCO configurations achieve zero violations of the min-max voltage and generation constraints, validating the physics-informed constraint handling. The unsupervised variants (PINCO-F-Phys, PINCO-U-Phys) deliver constraint violations comparable to IPOPT while achieving superior economic performance, with cost gaps ranging from -3.97\% to -1.82\% across test cases. Nevertheless, the supervised training (PINCO-F-Sup) tends to hinder the method's ability to respect constraints, particularly evident in the IEEE30 and Swiss test cases. This suggests that relying solely on IPOPT-generated labels may propagate solver-specific biases and constraint handling strategies that do not generalize well. In contrast, the unsupervised physics-informed approach (PINCO-F-Phys, PINCO-U-Phys) dynamically adapts its loss function to each constraint violation during training. It effectively implements a soft constraint handling mechanism that behaves similarly to hard constraint methods. Furthermore, incorporating semi-supervised learning with limited IPOPT feasibility labels (PINCO-U-Semi25) consistently improves agreement with IPOPT, as expected, yet the fully unsupervised variants demonstrate competitive performance independently, validating the robustness of the physics-informed learning paradigm. In \ref{app:semisupervised}, we analyze the impact of varying the number of feasibility labels.

Figure~\ref{fig:cost_gap_all_grids} presents instance-level analysis using only mutually feasible solutions from both methods. PINCO achieves lower costs in nearly all instances while maintaining constraint violations comparable to those of IPOPT, confirming its effectiveness as a fast, accurate alternative to traditional AC-OPF solvers. While PINCO exhibits higher maximum violations compared to IPOPT, primarily driven by power balance equality constraints, the violations remain within the same order of magnitude across most test instances. 
Higher violations are expected, given that PINCO employs a single model to predict solutions across all scenarios simultaneously, whereas IPOPT solves each instance individually with scenario-specific optimization. This architectural difference inherently introduces variance in per-instance performance but enables the dramatic computational advantages analyzed in Section~\ref {sec:time}.

The clustering branch in full PINCO models (PINCO-F variants) demonstrates strong feasibility classification performance, with agreement rates of 68-82\% against IPOPT. This metric reflects differences in the solution paths taken by the two methods rather than classification accuracy. PINCO sometimes identifies feasible solutions that IPOPT does not find, and vice versa, as both methods may converge to different local optima or explore different regions of the feasible space. As shown in Figure~\ref{fig:violation_comparison}, PINCO clearly separates feasible and infeasible clusters based on constraint violation patterns. Only power balance and thermal violations are displayed, as generation and voltage-bound violations are zero throughout. Importantly, the clustering branch learns these feasibility patterns during training by monitoring the global cluster structure across the entire dataset, thereby capturing the underlying manifold of constraint-violation distributions. This global perspective during training effectively translates to the test set, enabling clear separation of solutions and accurate assignment of feasibility labels, even for previously unseen operating conditions. While Figure~\ref{fig:violation_comparison} demonstrates clear separation between feasible and infeasible clusters within PINCO's framework, the divergence from IPOPT solutions highlights the inherent non-uniqueness of AC-OPF solutions and the different solution strategies employed by neural network and optimization-based approaches.

\begin{table}[htbp]
\centering
\caption{Computational comparison of PINCO and IPOPT across test
cases under sequential execution and best parallel configuration
(8 CPU workers for IPOPT; GPU batch size of 256 for PINCO).
Per-sample times are averaged over the test set.}
\label{tab:ipopt_pinco_comparison}
\resizebox{\columnwidth}{!}{%
\footnotesize
\begin{tabular}{llcccc}
\hline
& & \multicolumn{2}{c}{Sequential} & \multicolumn{2}{c}{Best Parallel} \\
\cmidrule(lr){3-4} \cmidrule(lr){5-6}
Case & Model
    & Total (s)$\downarrow$
    & Per sample (ms)$\downarrow$
    & Total (s)$\downarrow$
    & Per sample (ms)$\downarrow$ \\
\hline\hline
\multirow{2}{*}{IEEE30}
    & IPOPT & $20.41$ & $48.95$ & $4.28$  & $10.26$ \\
    & PINCO & $3.71$  & $6.34$  & $0.019$ & $0.032$ \\
\hline\hline
\multirow{2}{*}{IEEE57}
    & IPOPT & $17.72$ & $31.20$ & $3.66$  & $6.44$  \\
    & PINCO & $6.68$  & $6.02$  & $0.036$ & $0.033$ \\
\hline\hline
\multirow{2}{*}{Swiss}
    & IPOPT & $287.34$ & $118.29$ & $34.92$ & $14.37$ \\
    & PINCO & $18.14$  & $7.29$   & $0.127$ & $0.051$ \\
\hline
\end{tabular}%
}
\end{table}

\begin{figure}[h!]
    \centering
    \begin{subfigure}{\columnwidth}
        \centering
        \begin{overpic}[width=\columnwidth]{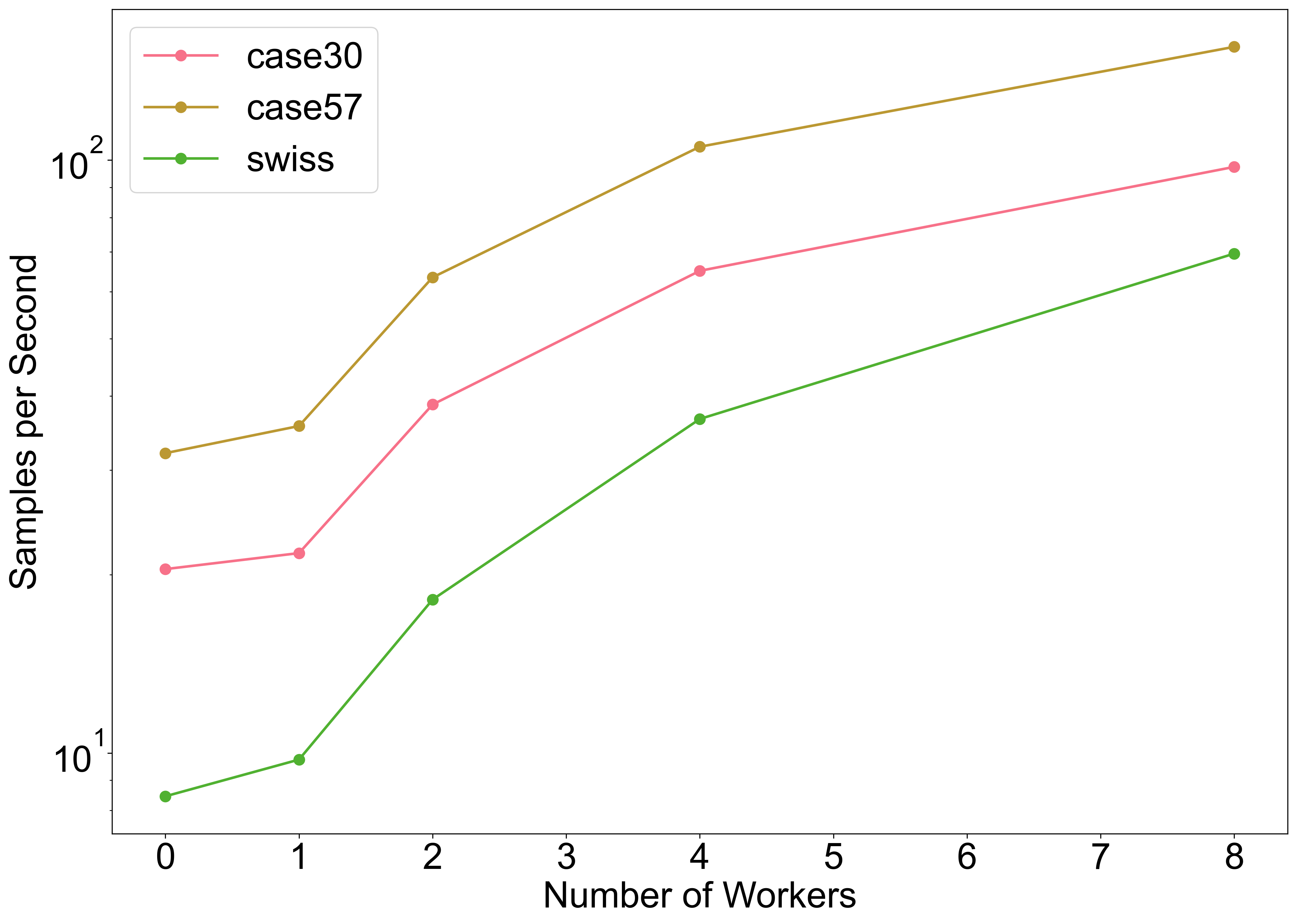}
            \put(1,1){(a)}
        \end{overpic}
        \phantomsubcaption
        \label{fig:ipopt_scaling}
    \end{subfigure}

    \vspace{1em}

    \begin{subfigure}{\columnwidth}
        \centering
        \begin{overpic}[width=\columnwidth]{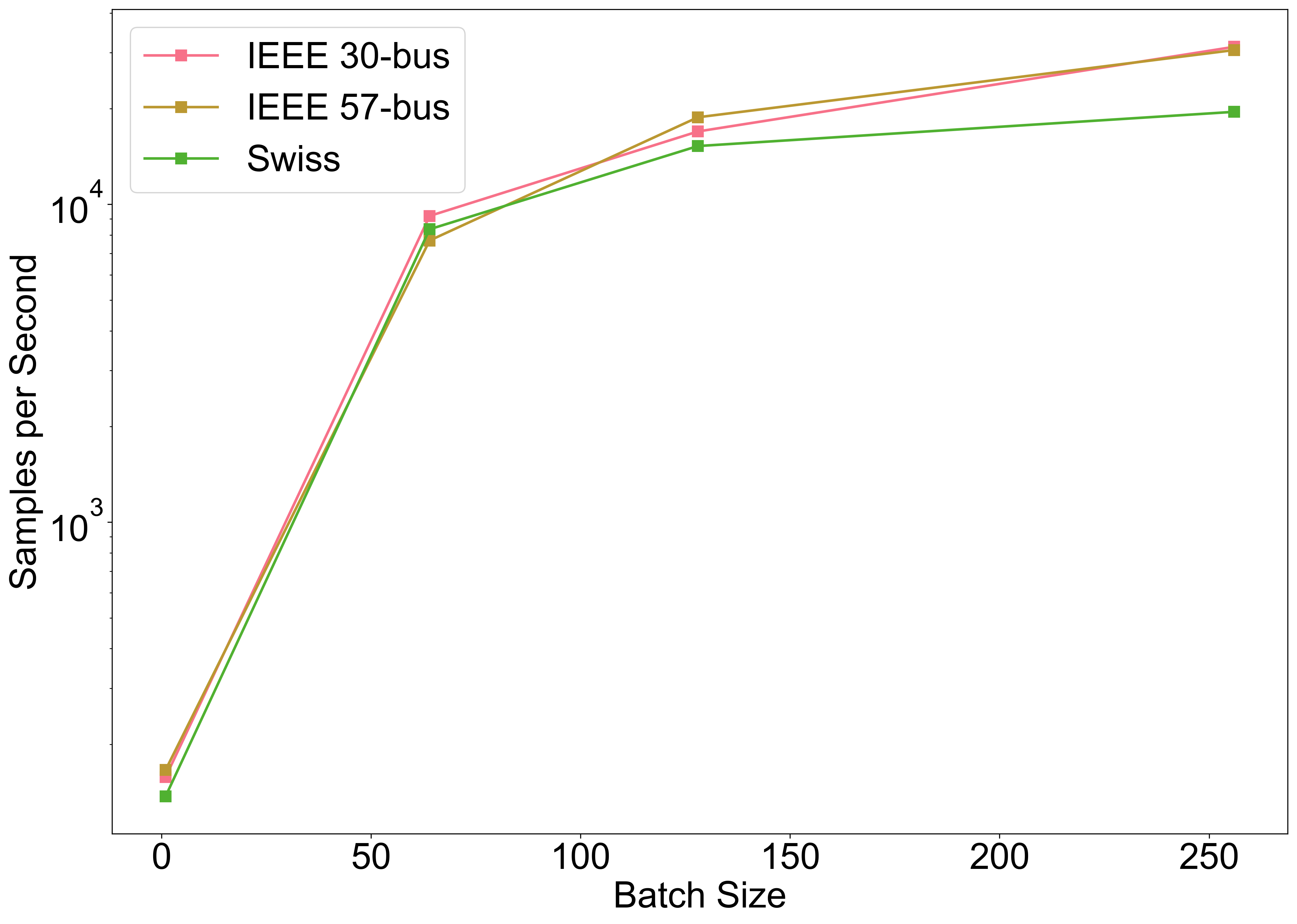}
            \put(1,1){(b)}
        \end{overpic}
        \phantomsubcaption
        \label{fig:pinco_scaling}
    \end{subfigure}

    \caption{Scaling performance of IPOPT~(a) and PINCO~(b) across
    IEEE 30-bus, IEEE 57-bus, and Swiss grid test cases. (a) IPOPT
    total computation time as a function of the number of parallel
    workers. (b) PINCO total computation time as a function of batch
    size.}
    \label{fig:scaling_comparison}
\end{figure}

\subsection{Computational Performance and Scaling Behavior}
\label{sec:time}
Comparing the computational speed of PINCO and IPOPT fairly is inherently challenging: the two methods are designed for different hardware and parallelisation paradigms. IPOPT is a sequential NLP solver that benefits from multi-core CPU parallelisation, while PINCO is a GPU-accelerated neural model whose throughput scales with batch size. In practice, both methods would be deployed in their respective optimal configurations; we, therefore, report results under both sequential execution and best parallel configuration for each method, and encourage the reader to interpret the comparison with this asymmetry in mind. Table~\ref{tab:ipopt_pinco_comparison} reports total and per-sample inference times under sequential execution and under the best parallel configuration for each method (8 CPU workers for IPOPT; GPU batch processing for PINCO). Under sequential execution, PINCO already achieves substantially lower per-sample times than IPOPT across all test cases. The scaling analysis in Figure~\ref{fig:scaling_comparison} further reveals fundamentally different performance characteristics: IPOPT achieves at most $2$--$3\times$ speedup with up to 8 workers, while PINCO's throughput scales rapidly with batch size before saturating at batch sizes of $50$--$100$, maintaining a consistent throughput of $10^3$--$10^4$ samples per second beyond that point. This behaviour suggests that PINCO is particularly well suited for scenarios requiring repeated solves over large instance sets, such as real-time contingency screening or scenario-based planning, where batch inference over GPU hardware provides a structural advantage over sequential NLP solving.

\section{Limitations} \label{sec:limitations}
While PINCO demonstrates significant computational advantages and competitive solution quality compared to traditional optimization methods, some limitations warrant discussion.
First, PINCO requires substantial upfront investment in training time. The GNN architecture, once determined, remains consistent across all test cases, requiring no grid-specific tuning. However, the training process itself is computationally intensive, requiring convergence of the physics-informed loss function before the model can be deployed. Once trained, the model generalises across varying operating conditions and topologies within the same grid without retraining. A limitation concerns the scope and scalability of the training dataset. Our evaluation considers contingency scenarios up to N-2 outages, which, while comprehensive for many practical applications, represents a subset of possible grid configurations. Extending the framework to higher-order outages (N-3 and beyond), combined with diverse loading scenarios, would create exponentially larger datasets that become computationally intractable to generate and store. This constraint limits our ability to train models that cover the full space of rare but potentially critical grid states. Consequently, the method's performance on extreme or previously unseen outage combinations remains an open question. Future work should focus on extending this method to comprehensive benchmark datasets. Another potential limitation is generalisation across different grids: a model trained on one network may not directly transfer to a structurally different network, requiring a dedicated training run for each new grid.

\section{Conclusions}
We have presented PINCO, an unsupervised physics-informed learning framework that addresses fundamental challenges in applying deep learning to AC optimal power flow problems. By embedding power flow physics directly into the loss function and introducing a clustering-based feasibility classification mechanism, PINCO eliminates the need for pre-screening feasible scenarios, a critical limitation of previous approaches. Our comprehensive evaluation across multiple test cases demonstrates that PINCO achieves computational speedups over IPOPT and DeepOPF-FT. Moreover, it maintains constraint violations comparable to those of traditional solvers and cheaper solutions, with cost reductions of 0.5-4\%. The framework's consistent architecture across different grid topologies, from IEEE test cases to the Swiss power system, validates its generalizability and practical applicability. While limitations remain, PINCO establishes a compelling pathway toward real-time grid optimization in increasingly complex power systems. The combination of unsupervised learning, physics-informed constraints, and feasibility detection positions this approach as a viable complement to traditional optimization methods, particularly in scenarios requiring rapid decision-making under uncertainty. Future work will focus on extending the framework to larger-scale grids, incorporating dynamic operating conditions, and investigating hybrid approaches that combine neural network speed with strict feasibility enforcement for safety-critical applications.
\label{sec:conclusions}





\bibliographystyle{elsarticle-num-names}
\bibliography{refs}

@inproceedings{mohammadian2025restoring,
  author    = {Mohammadian, Mostafa and Van Boven, Anna and Baker, Kyri},
  title     = {Restoring Feasibility in Power Grid Optimization:
               A Counterfactual {ML} Approach},
  booktitle = {2025 IEEE International Conference on Communications,
               Control, and Computing Technologies for Smart Grids
               (SmartGridComm)},
  year      = {2025},
  doi       = {10.1109/SmartGridComm65349.2025.11204567}
}

@ARTICLE{topaware,
  author={Liu, Shaohui and Wu, Chengyang and Zhu, Hao},
  journal={IEEE Transactions on Power Systems}, 
  title={Topology-Aware Graph Neural Networks for Learning Feasible and Adaptive AC-OPF Solutions}, 
  year={2023},
  volume={38},
  number={6},
  pages={5660-5670},
  doi={10.1109/TPWRS.2022.3230555}}

@misc{MLOPFWiki2025,
  title = {ML OPF wiki: Machine Learning for Solving Optimal Power Flow Problems},
  author = {Chen, Minghua and Pan, Xiang and Zhao, Jiawei and Zhou, Min},
  year = {2025},
  month = {December},
  url = {https://energy.hosting.acm.org/wiki/index.php/ML_OPF_wiki},
  note = {Accessed online}
}

@ARTICLE{MATPOWER,
  author={Zimmerman, Ray Daniel and Murillo-Sánchez, Carlos Edmundo and Thomas, Robert John},
  journal={IEEE Transactions on Power Systems}, 
  title={MATPOWER: Steady-State Operations, Planning, and Analysis Tools for Power Systems Research and Education}, 
  year={2011},
  volume={26},
  number={1},
  pages={12-19},
  doi={10.1109/TPWRS.2010.2051168}
}

@article{kile2014,
  title={{A comparison of AC and DC power flow models for contingency and reliability analysis}},
  author={Hakon Kile and Kjetil Uhlen and Leif Warland and Gerd H. Kj{\o}lle},
  journal={2014 Power Systems Computation Conference},
  year={2014},
  pages={1-7}
}

@article{deepopf,
  title={{DeepOPF: A Feasibility-Optimized Deep Neural Network Approach for AC Optimal Power Flow Problems}},
  author={Pan, Xiang and Zhao, Tianyu and Chen, Minghua},
  journal={IEEE Transactions on Smart Grid},
  volume={11},
  number={6},
  pages={4883--4893},
  year={2020},
  publisher={IEEE}
}

@ARTICLE{deepopf-ft,
  author={Zhou, Min and Chen, Minghua and Low, Steven H.},
  journal={IEEE Transactions on Power Systems}, 
  title={{DeepOPF-FT: One Deep Neural Network for Multiple AC-OPF Problems With Flexible Topology}}, 
  year={2023},
  volume={38},
  number={1},
  pages={964-967},
  doi={10.1109/TPWRS.2022.3217407}}

@ARTICLE{IPM,
  author={Wang, Hongye and Murillo-Sanchez, Carlos E. and Zimmerman, Ray D. and Thomas, Robert J.},
  journal={IEEE Transactions on Power Systems}, 
  title={On Computational Issues of Market-Based Optimal Power Flow}, 
  year={2007},
  volume={22},
  number={3},
  pages={1185-1193},
  doi={10.1109/TPWRS.2007.901301}}

@article{lu2021physics,
  title={Physics-informed neural networks with hard constraints for inverse design},
  author={Lu, Lu and Pestourie, Raphael and Yao, Wenjie and Wang, Zhicheng and Verdugo, Francesc and Johnson, Steven G},
  journal={SIAM Journal on Scientific Computing},
  volume={43},
  number={6},
  pages={B1105--B1132},
  year={2021},
  publisher={SIAM}
}

@inproceedings{owerko2022unsupervisedoptimalpowerflow,
  title={Unsupervised optimal power flow using graph neural networks},
  author={Owerko, Damian and Gama, Fernando and Ribeiro, Alejandro},
  booktitle={ICASSP 2024-2024 IEEE International Conference on Acoustics, Speech and Signal Processing (ICASSP)},
  pages={6885--6889},
  year={2024},
  organization={IEEE}
}

@inproceedings{DCvsAC,
  title={{Solutions of DC OPF are never AC feasible}},
  author={Baker, Kyri},
  booktitle={Proceedings of the Twelfth ACM International Conference on Future Energy Systems},
  pages={264--268},
  year={2021}
}

@incollection{opftheory,
title = {Mathematical programming for power systems},
editor = {Jorge García},
booktitle = {Encyclopedia of Electrical and Electronic Power Engineering},
publisher = {Elsevier},
address = {Oxford},
pages = {722-733},
year = {2023},
isbn = {978-0-12-823211-8},
doi = {https://doi.org/10.1016/B978-0-12-821204-2.00044-1},
url = {https://www.sciencedirect.com/science/article/pii/B9780128212042000441},
author = {Salvador Pineda and Juan Miguel Morales and Sonja Wogrin},
}

@article{Nair2024ACOPF,
  author = {Arun Sukumaran Nair and Shrirang Abhyankar and Slaven Peles and Prakash Ranganathan},
  title = {{Computational and Numerical Analysis of AC Optimal Power Flow Formulations on Large-Scale Power Grids}},
  journal = {IET Research Journals},
  year = {2024},
  volume = {TBD},
  number = {TBD},
  pages = {TBD},
  doi = {0000000000},
  issn = {1751-8644},
  url = {www.ietdl.org},
  publisher = {Institution of Engineering and Technology},
  address = {Grand Forks, ND, USA and Richland, WA, USA}
}

@INPROCEEDINGS{convexACOPF,
  author={Lavaei, Javad and Low, Steven H.},
  booktitle={2010 48th Annual Allerton Conference on Communication, Control, and Computing (Allerton)}, 
  title={Convexification of optimal power flow problem}, 
  year={2010},
  volume={},
  number={},
  pages={223-232},
  doi={10.1109/ALLERTON.2010.5706911}}

@misc{canos,
      title={{CANOS: A Fast and Scalable Neural AC-OPF Solver Robust To N-1 Perturbations}}, 
      author={Luis Piloto and Sofia Liguori and Sephora Madjiheurem and Miha Zgubic and Sean Lovett and Hamish Tomlinson and Sophie Elster and Chris Apps and Sims Witherspoon},
      year={2024},
      eprint={2403.17660},
      archivePrefix={arXiv},
      primaryClass={cs.LG},
      url={https://arxiv.org/abs/2403.17660}, 
}

@article{raissi2019physics,
  title={Physics-informed neural networks: A deep learning framework for solving forward and inverse problems involving nonlinear partial differential equations},
  author={Raissi, Maziar and Perdikaris, Paris and Karniadakis, George E},
  journal={Journal of Computational physics},
  volume={378},
  pages={686--707},
  year={2019},
  publisher={Elsevier}
}

@ARTICLE{opf_relaxation_1,
    author = {S.H. Low},
    journal = {IEEE Transactions on Control of Network Systems},
    title = {{Convex Relaxation of Optimal Power Flow—Part II: Exactness}},
    year = {2014},
    volume = {1},
    number = {2},
    pages = {177-189},
    doi = {10.1109/TCNS.2014.2323634},
}

@manual{zimmerman2016matpower,
  title={Matpower Interior Point Solver MIPS 1.3 User's Manual},
  author={R. D. Zimmerman and H. Wang},
  year={2016}
}

@misc{ieee30bus,
  title        = {University of Washington Power Systems Test Case Archive: IEEE 30-Bus Power Flow Test Case},
  author       = {Rich Christie},
  howpublished = {\url{http://labs.ece.uw.edu/pstca/pf30/pg\_tca30bus.htm}},
  note         = {Accessed: 2024-09-22},
  year         = {1993},
  institution  = {University of Washington}
}

@article{Bienstock2022,
  title={Mathematical programming formulations for the alternating current optimal power flow problem},
  author={Bienstock, Daniel and Escobar, Mauro and Gentile, Claudio and Liberti, Leo},
  journal={Annals of Operations Research},
  volume={314},
  number={1},
  pages={277--315},
  year={2022},
  publisher={Springer},
  doi={10.1007/s10479-021-04497-z},
  url={https://doi.org/10.1007/s10479-021-04497-z},
  issn={1572-9338}
}

@article{VARBELLA2023,
title = {Geometric deep learning for online prediction of cascading failures in power grids},
journal = {Reliability Engineering \& System Safety},
volume = {237},
pages = {109341},
year = {2023},
issn = {0951-8320},
doi = {https://doi.org/10.1016/j.ress.2023.109341},
url = {https://www.sciencedirect.com/science/article/pii/S0951832023002557},
author = {Anna Varbella and Blazhe Gjorgiev and Giovanni Sansavini},
}

@ARTICLE{piGnnpert,
  author={Yang, Mei and Qiu, Gao and Liu, Junyong and Liu, Youbo and Liu, Tingjian and Tang, Zhiyuan and Ding, Lijie and Shui, Yue and Liu, Kai},
  journal={IEEE Transactions on Industrial Informatics}, 
  title={Topology-Transferable Physics-Guided Graph Neural Network for Real-Time Optimal Power Flow}, 
  year={2024},
  volume={20},
  number={9},
  pages={10857-10872},
  doi={10.1109/TII.2024.3398058}}

@misc{eulerwiki,
  title = "Euler Wiki",
  author = "Swiss National Supercomputing Centre (CSCS)",
  year = "2023",
  howpublished = "\url{https://scicomp.ethz.ch/wiki/Euler}",
  note = "[Accessed: April 26, 2023]"
}

@article{lagrangian-duality,
  title={{Unsupervised Deep Learning for AC Optimal Power Flow via Lagrangian Duality}},
  author={Chen, Yize and Wang, Xiyang and Zheng, Baosen},
  journal={IEEE Transactions on Power Systems},
  volume={37},
  number={5},
  pages={3982--3988},
  year={2022},
  publisher={IEEE}
}

@inproceedings{selfsupervised,
  title={Self-supervised primal-dual learning for constrained optimization},
  author={Park, Seonho and Van Hentenryck, Pascal},
  booktitle={Proceedings of the AAAI Conference on Artificial Intelligence},
  volume={37},
  number={4},
  pages={4052--4060},
  year={2023}
}

@article{lagrangian-approach,
  title={{Learning to Solve the AC Optimal Power Flow via a Lagrangian Approach}},
  author={Jin, Chen and Chen, Yize and Xu, Qixuan and Zeng, Xiang and Huang, Lei and Lang, Yubo},
  journal={IEEE Transactions on Smart Grid},
  volume={12},
  number={5},
  pages={4008--4019},
  year={2021},
  publisher={IEEE}
}

@inproceedings{dc3,
  title={{DC3: A learning method for optimization with hard constraints}},
  author={Donti, Priya L and Rolnick, David and Kolter, J Zico},
  booktitle={International Conference on Learning Representations},
  url={https://openreview.net/forum?id=V1ZHVxJ6dSS}
}

@misc{swissgrid,
    author = {SwissGrid},
    title = {www.swissgrid.ch},
    url   = {https://www.swissgrid.ch/en/home/operation/grid-data/transmission.html#downloads},
    year={2009},
    addendum = {(accessed: 23.12.2022)},
}

@inproceedings{TransformerConv,
  title={{Masked Label Prediction: Unified Message Passing Model for Semi-Supervised Classification}},
  author={Shi, Yunsheng and Huang, Zhengjie and Feng, Shikun and Zhong, Hui and Wang, Wenjing and Sun, Yu},
  booktitle={Proceedings of the Thirtieth International Joint Conference on Artificial Intelligence},
  pages={1548--1554},
  year={2021},
  organization={International Joint Conferences on Artificial Intelligence Organization}
}

@misc{ieee57bus,
  author = {{University of Washington}},
  title = {{Power Systems Test Case Archive: 57 Bus Power Flow Test Case}},
  howpublished = {\url{https://labs.ece.uw.edu/pstca/pf57/pg_tca57bus.htm}},
  year={1993},
  note = {Accessed: 2025}
}

@article{wang2022and,
  title={{When and why PINNs fail to train: A neural tangent kernel perspective}},
  author={Wang, Sifan and Yu, Xinling and Perdikaris, Paris},
  journal={Journal of Computational Physics},
  volume={449},
  pages={110768},
  year={2022},
  publisher={Elsevier}
}

@article{bose2025presolving,
  title={Presolving convexified optimal power flow with mixtures of gradient experts},
  author={Bose, Shourya and Chen, Kejun and Zhang, Yu},
  journal={Energy and AI},
  volume={22},
  pages={100417},
  year={2025},
  publisher={Elsevier},
  doi={10.1016/j.egyai.2025.100417}
}

@article{wolgast2024learning,
  title={Learning the optimal power flow: Environment design matters},
  author={Wolgast, Thomas and Nie{\ss}e, Astrid},
  journal={Energy and AI},
  volume={18},
  pages={100410},
  year={2024},
  publisher={Elsevier},
  doi={10.1016/j.egyai.2024.100410}
}

@article{khaloie2025review,
  title={Review of machine learning techniques for optimal power flow},
  author={Khaloie, Hamid and Dolanyi, Marin and Toubeau, Jean-Fran{\c{c}}ois and Vall{\'e}e, Fran{\c{c}}ois},
  journal={Applied Energy},
  volume={388},
  pages={125637},
  year={2025},
  publisher={Elsevier},
  doi={10.1016/j.apenergy.2025.125637}
}

@article{nguyen2025fsnet,
  title={FSNet: Feasibility-seeking neural network for constrained optimization with guarantees},
  author={Nguyen, Hoang and Donti, Priya},
  journal={Advances in Neural Information Processing Systems},
  volume={38},
  pages={39670--39708},
  year={2025}
}
\onecolumn
\clearpage
\twocolumn

\appendix

\section{Augmented Lagrangian Training}
\label{apx:al}

The physics loss is formulated using an Augmented Lagrangian (AL) approach. 
At each training step, the constraint violation-related loss is:

\begin{equation}
\mathcal{L} = \bar{\lambda}_{eq} \cdot \mathcal{L}_{eq} + \bar{\lambda}_{flow} \cdot \mathcal{L}_{flow} 
\end{equation}
where $\bar{\lambda}_{eq}$ and $\bar{\lambda}_{flow}$ are per-sample Lagrange multiplier vectors averaged over the batch, $\mathcal{L}_{eq}$ is the power balance violation, $\mathcal{L}_{flow}$ is the thermal flow violation. Generator and voltage inequality constraints are enforced implicitly through the network's output-layer bounds and are not included in the AL terms.
The multipliers are maintained per training sample and updated at the end of each epoch as:

\begin{equation}
\lambda_{eq}^{(k+1)} \leftarrow \lambda_{eq}^{(k)} + 2\mu_f \cdot \mathcal{L}_{eq}^{(k)}
\end{equation}
\begin{equation}
\lambda_{flow}^{(k+1)} \leftarrow \lambda_{flow}^{(k)} + 2\mu_h \cdot \mathcal{L}_{flow}^{(k)}
\end{equation}
where $\mu_f = 0.1$ and $\mu_h = 0.1$ are fixed step sizes and $k$ denotes the epoch index.

\section{Hyperparameters}
\label{apx:hyp}
We report here all the hyperparameters used for the PINCO experiments across the three test cases: IEEE 30-bus, IEEE 57-bus, and the Swiss network. For all IPOPT solves, we use the default constraint violation tolerance \texttt{constr\_viol\_tol}$=10^{-4}$~p.u., corresponding to $10^{-2}$~MVA on a 100~MVA base; consequently, the constraint violations reported in the main text are expressed in MVA, while IPOPT's convergence criterion is defined in per-unit values. The architecture consists of a network with $N_{mpl}$ message-passing layers~\cite{TransformerConv}, followed by two linear layers with tanh activations. The same hyperparameters are used across all three test cases. 
\begin{table}[h]
\centering
\caption{Hyperparameters for PINCO experiments on the IEEE 30-bus, IEEE 57-bus, and Swissgrid test cases.}
\label{tab:hyperparameters}
\footnotesize
\begin{tabular}{ll}
\hline
\textbf{Hyperparameter} & \textbf{Value} \\ \hline
Hidden dimension        & 64 \\
$N_{mpl}$              & 5 \\
Attention heads         & 1 \\
Activation function     & tanh \\
Batch size              & 256 \\
Learning rate           & 0.001 \\
$\gamma$               & 0.9995 \\
\end{tabular}
\end{table}
The learning rate decays exponentially by a factor $\gamma$ every 10 epochs.
\paragraph{DeepOPF-ft hyperparameter}
For the DeepOPF-FT comparison, the baseline model is implemented as a fully-connected network with four hidden layers of sizes 1024, 512, 256, and 113 units, respectively, followed by a sigmoid output layer. The model is trained for 5000 epochs using the Adam optimizer with an initial learning rate of $10^{-3}$, a step learning rate scheduler with decay factor $\gamma = 0.9$ applied every 100 epochs, and a batch size of 128.

\section{Semi-Supervised Learning Results}
\label{app:semisupervised}

This appendix presents detailed results for the semi-supervised learning variants of PINCO across all three test cases. We evaluate PINCO-U-Semi1, PINCO-U-Semi5, and PINCO-U-Semi10, trained with 1\%, 5\%, and 10\% of binary labels, respectively, on sample feasibility.

\begin{table}[htpb]
\centering
\caption{Semi-supervised learning results across test cases}
\label{tab:semisupervised_appendix}
\scriptsize
\setlength{\tabcolsep}{3pt}
\begin{tabular}{llccc}
\hline
Model & Metric & IEEE30 & IEEE57 & Swiss \\
\hline
\multirow{5}{*}{PINCO-U-Semi1} 
& Power Bal.$\downarrow$ & $4.31\cdot10^{-1}$ & $1.32$ & $3.16$ \\
& Thermal$\downarrow$ & $\mathbf{0}$ & $\mathbf{0}$ & $\mathbf{0}$ \\
& Cost$\downarrow$ & $5.71\cdot10^{2}$ & $\mathbf{3.97\cdot10^{4}}$ & $7.01\cdot10^{4}$ \\
& Gap\%$\downarrow$ & $-1.74$ & $\mathbf{-5.13}$ & $\mathbf{-0.25}$ \\
& Agreement\% & $74.36$ & $62.70$ & $65.90$ \\
\hline
\multirow{5}{*}{PINCO-U-Semi5} 
& Power Bal.$\downarrow$ & $4.38\cdot10^{-1}$ & $\mathbf{1.19}$ & $5.58$ \\
& Thermal$\downarrow$ & $1.21\cdot10^{-3}$ & $\mathbf{0}$ & $\mathbf{0}$ \\
& Cost$\downarrow$ & $5.73\cdot10^{2}$ & $4.02\cdot10^{4}$ & $7.51\cdot10^{4}$ \\
& Gap\%$\downarrow$ & $-1.31$ & $-3.94$ & $6.91$ \\
& Agreement\% & $\mathbf{78.97}$ & $56.49$ & $\mathbf{72.41}$ \\
\hline
\multirow{5}{*}{PINCO-U-Semi10} 
& Power Bal.$\downarrow$ & $7.00\cdot10^{-1}$ & $1.32$ & $\mathbf{3.00}$ \\
& Thermal$\downarrow$ & $1.75\cdot10^{-4}$ & $\mathbf{0}$ & $\mathbf{0}$ \\
& Cost$\downarrow$ & $5.88\cdot10^{2}$ & $3.96\cdot10^{4}$ & $\mathbf{6.98\cdot10^{4}}$ \\
& Gap\%$\downarrow$ & $\mathbf{1.27}$ & $-5.32$ & $-0.67$ \\
& Agreement\% & $70.26$ & $\mathbf{65.68}$ & $74.30$ \\
\hline
\end{tabular}
\end{table}

\section{Robustness Across Random Seeds}
\label{app:seeds}

We evaluate \mbox{PINCO-F-Phys} and \mbox{PINCO-U-Phys} across three random seeds on the IEEE 30-bus, IEEE 57-bus, and Swiss power grid test cases. We report the two representative configurations described in Section \ref{sec:datasetpinco}.
Supervised and semi-supervised baselines are omitted as a full ablation replication across seeds was precluded by computational constraints. Bound inequality constraints are satisfied by construction and are
therefore excluded from this analysis. Table~\ref{tab:seed_robustness} reports mean~$\pm$~std over seeds for power balance violation, generation cost, and cost gap.

\begin{table}[h]
\centering
\caption{Robustness evaluation of \mbox{PINCO-F-Phys} and
\mbox{PINCO-U-Phys} across multiple random seeds (mean $\pm$ std).
Power balance and thermal violation in MVA; cost in \$.
Bound inequality violations are zero by construction and omitted.
All reported cases are feasible.}
\label{tab:seed_robustness}
\scriptsize
\setlength{\tabcolsep}{4pt}
\begin{tabular}{llcc}
\hline
Case & Metric & PINCO-F-Phys & PINCO-U-Phys \\
\hline\hline
\multirow{3}{*}{IEEE30}
 & Power Bal.\ (MVA) & $4.74{\cdot}10^{-1} \pm 2.10{\cdot}10^{-2}$ & $5.08{\cdot}10^{-1} \pm 3.12{\cdot}10^{-2}$ \\
 & Thermal (MVA)     & $2.12{\cdot}10^{-3} \pm 1.96{\cdot}10^{-4}$ & $1.85{\cdot}10^{-3} \pm 1.98{\cdot}10^{-4}$ \\
 & Cost (\$)         & $5.90{\cdot}10^{2} \pm 1.52{\cdot}10^{1}$   & $5.71{\cdot}10^{2} \pm 4.29{\cdot}10^{0}$   \\
\hline\hline
\multirow{3}{*}{IEEE57}
 & Power Bal.\ (MVA) & $1.05{\cdot}10^{0} \pm 2.59{\cdot}10^{-1}$ & $1.15{\cdot}10^{0} \pm 1.92{\cdot}10^{-1}$ \\
 & Thermal (MVA)     & $0.00 \pm 0.00$                             & $0.00 \pm 0.00$                             \\
 & Cost (\$)         & $4.05{\cdot}10^{4} \pm 5.34{\cdot}10^{2}$  & $4.00{\cdot}10^{4} \pm 5.28{\cdot}10^{2}$  \\
\hline\hline
\multirow{3}{*}{Swiss}
 & Power Bal.\ (MVA) & $ 2.88{\cdot}10^{0} \pm 5.62{\cdot}10^{-1} $ & $2.61{\cdot}10^{0} \pm 9.66{\cdot}10^{-2}$ \\
 & Thermal (MVA)     & $0.00 \pm 0.00$   & $0.00 \pm 0.00$  \\
 & Cost (\$)         & $6.88{\cdot}10^{4} \pm 5.75{\cdot}10^{2}$ & $6.91{\cdot}10^{4} \pm 1.25{\cdot}10^{2}$  \\
\hline
\end{tabular}
\end{table}
\end{document}